\documentclass[twocolumn,aps,nofootinbib]{revtex4-1}
\usepackage{amsfonts, amssymb}
\usepackage{graphicx, epsfig,bm}
\usepackage{color}

\usepackage{amsbsy}
\usepackage{amsfonts}
\usepackage{threeparttable}
\usepackage{amsmath}
\usepackage{amssymb}

\newcommand{\be}{\begin{equation}}
\newcommand{\ee}{\end{equation}}
\newcommand{\bea}{\begin{eqnarray}}
\newcommand{\eea}{\end{eqnarray}}

\newcommand{\PRL}{{Phys. Rev. Lett.\,}}

\def\fun#1#2{\lower3.6pt\vbox{\baselineskip0pt\lineskip.9pt
        \ialign{$\mathsurround=0pt#1\hfill##\hfil$\crcr#2\crcr\sim\crcr}}}

\newcommand{\neff}{\ensuremath{N_\mathrm{eff}}}
\newcommand{\sumnu}{\ensuremath{\sum{m_{\nu}}}}
\newcommand{\yhe}{\ensuremath{Y_p}}
\newcommand{\nrun}{\ensuremath{dn_s/d\ln k}}

\newcommand\lsim{\mathrel{\rlap{\lower4pt\hbox{\hskip1pt$\sim$}}
    \raise1pt\hbox{$<$}}}
\newcommand\gsim{\mathrel{\rlap{\lower4pt\hbox{\hskip1pt$\sim$}}
    \raise1pt\hbox{$>$}}}

\def\dslash{\not{\hbox{\kern-2pt $\partial$}}}
\def\Dslash{\not{\hbox{\kern-4pt $D$}}}
\def\Oslash{\not{\hbox{\kern-4pt $O$}}}
\def\Qslash{\not{\hbox{\kern-4pt $Q$}}}
\def\pslash{\not{\hbox{\kern-2.3pt $p$}}}
\def\kslash{\not{\hbox{\kern-2.3pt $k$}}}
\def\qslash{\not{\hbox{\kern-2.3pt $q$}}}

 \newtoks\slashfraction
 \slashfraction={.13}
 \def\slash#1{\setbox0\hbox{$ #1 $}
 \setbox0\hbox to \the\slashfraction\wd0{\hss \box0}/\box0 }

\def\ee{\end{equation}}
\def\be{\begin{equation}}

\newcommand\Tr{{\rm Tr}\,}

\begin{document}
\setlength{\unitlength}{1mm}
\title{Constraints on Neutrino Mass and Light Degrees of Freedom \\ in Extended Cosmological Parameter Spaces}  
\author{Shahab Joudaki}%\email{joudaki@uci.edu}
\affiliation{Center for Cosmology, Dept.~of Physics \& Astronomy, University of California, Irvine, CA 92697}

\date{\today}

\begin{abstract}
From a combination of probes including the cosmic microwave background (WMAP7+SPT), Hubble constant (HST), baryon acoustic oscillations (SDSS+2dFGRS), and supernova distances (Union2), 
we have explored the extent to which the constraints on the effective number of neutrinos and sum of neutrino masses are affected by our ignorance of other cosmological parameters, including the curvature of the universe, running of the spectral index, primordial helium abundance, evolving late-time dark energy, and early dark energy. 
In a combined analysis of the effective number of neutrinos and sum of neutrino masses, we find mild (2.2$\sigma$) preference for additional light degrees of freedom. However, the effective number of neutrinos is consistent with the canonical expectation of 3 massive neutrinos and no extra relativistic species to within $1\sigma$ when allowing for evolving dark energy and relaxing the strong inflation prior on the curvature and running. The agreement improves with the possibility of an early dark energy component, itself constrained to be less than 5\% of the critical density (95\% CL) in our expanded parameter space.
In extensions of the standard cosmological model, the derived amplitude of linear matter fluctuations $\sigma_8$ is found to closely agree with low-redshift cluster abundance measurements.
The sum of neutrino masses is robust to assumptions of the effective number of neutrinos, late-time dark energy, curvature, and running at the level of 1.2 eV (95\% CL). 
The upper bound degrades to 2.0 eV (95\% CL) when further including the early dark energy density and primordial helium abundance as additional free parameters.
Even in extended cosmological parameter spaces, Planck alone could determine the possible existence of extra relativistic species at $4\sigma$ confidence and constrain the sum of neutrino masses to 0.2 eV (68\%~CL).

\end{abstract}
\bigskip

\maketitle

\section{Introduction}
\label{introlabel}

Observations of the cosmic microwave background (CMB)~\cite{Spergel:2003cb,Spergel:2006hy,Komatsu,Komatsu:2010fb}, large-scale structure~\cite{Tegmark:2003uf,Tegmark:2006az,Cole:2005sx}, and type Ia supernovae (SNe)~\cite{RiessSNe,Perlmutter} have established a flat $\Lambda$CDM model, with nearly scale-invariant, adiabatic, Gaussian primordial fluctuations as providing a consistent description of the global properties of our universe.
At the same time, we do not yet understand the microscopic identities of the dark energy ($\Lambda$), cold dark matter (CDM), and inflaton (primordial fluctuations) that enter our standard cosmological model.

\begin{table}[t!]
\vspace{0.5em}
\begin{center}
\begin{tabular}{lc|c}
\hline\hline
Parameter & Symbol & Prior\\
\hline
Baryon density & $\Omega_{b}h^2$ & $0.005 \to 0.1$\\
Dark matter density & $\Omega_{c}h^2$ & $0.01 \to 0.99$\\
Angular size of sound horizon & $\theta_s$ & $0.5 \to 10$\\
Optical depth to reionization & $\tau$ & $0.01 \to 0.8$\\
Scalar spectral index & $n_{s}$ & $0.5 \to 1.5$\\
Amplitude of scalar spectrum & $\ln{(10^{10} A_{s})}$ & $2.7 \to 4$\\
\hline
Effective number of neutrinos & \neff &  $1.047 \to 10$\\
Sum of neutrino masses & $\sum{m_{\nu}}~\rm{[eV]}$ &  $0 \to 5$\\
Constant dark energy EOS & $w$ & $-3 \to 0$\\
Running of the spectral index & ${dn_s \over d\ln k}$ &  $-0.2 \to 0.2$\\
Curvature of the universe & $\Omega_{k}$ &  $-0.4 \to 0.4$\\
Primordial helium abundance & $Y_p$ &  $0 \to  1$\\
\hline
Present dark energy EOS & $w_0$ & $-3 \to 0$\\
Derivative of dark energy EOS & $w_a$ &  $-10 \to 10$\\
Early dark energy density & $\Omega_e$ & $0 \to  0.2$\\
\hline\hline
\end{tabular}
\caption{We impose uniform priors on the above cosmological parameters. 
In addition, we always consider the Poisson point source power $D_{3000}^{\rm{PS}}$, the clustered power $D_{3000}^{\rm{CL}}$, and the SZ power $D_{3000}^{\rm{SZ}}$ as nuisance parameters constrained by the CMB data~\cite{Keisler:2011aw}.
Moreover, we always derive $\sigma_8$, the amplitude of linear matter fluctuations on scales of $8~{\rm{Mpc}}/h$ at $z=0$.
Beyond a constant dark energy equation of state (EOS), we also consider a time-varying expansion $w(a) = w_0 + (1-a)w_a$, and an early dark energy model described in Sec.~\ref{subformafive}.
In this table, the first 6 parameters are defined as ``vanilla" parameters.}
\vspace{-2.5em}
\label{table:priors}
\end{center}
\end{table}

\begin{table*}[t!]\footnotesize
\begin{center}
\caption{Constraints on Cosmological Parameters using SPT+WMAP+$H_0$+BAO.}
\begin{tabular}{lcccccccccccccc|}
\hline \hline
 & & $\Lambda$CDM & $\Lambda$CDM & $\Lambda$CDM & $\Lambda$CDM & $\Lambda$CDM & $\Lambda$CDM\\
 & & & +~\neff & +~$\sum{m_\nu}$ & +\neff+$\sum{m_\nu}$ & +\neff+$Y_p$ & +\neff+$\sum{m_\nu}$+$Y_p$\\
\hline \hline
Primary & $100 \Omega_b h^2$ & $2.237 \pm 0.038$  & $2.261 \pm 0.042$ &  $2.238 \pm 0.039$  & $2.272 \pm 0.043$ & $2.271 \pm 0.043$ & $2.273 \pm 0.044$ \cr
& $100\Omega_c h^2$ & $11.22 \pm 0.28$  & $12.80 \pm 0.92$ & $11.11 \pm 0.29$    & $13.05 \pm 0.94$ & $12.5 \pm 1.1$ & $13.0 \pm 1.2$ \cr
& $10^4\theta_s$ & $104.12 \pm 0.15$ & $103.95 \pm 0.17$  & $104.15 \pm 0.15$  & $103.95 \pm 0.18$ & $104.07 \pm 0.29$ & $103.96 \pm 0.29$ \cr
& $\tau$ & $0.086 \pm 0.014$ & $0.086 \pm 0.014$ & $0.088 \pm 0.014$ & $0.090 \pm 0.015$ & $0.088 \pm 0.014$ & $0.090 \pm 0.015$ \cr
& $n_s$ & $0.9648 \pm 0.0092$ & $0.981 \pm 0.013$ & $0.9661 \pm 0.0096$ & $0.987 \pm 0.013$ & $0.983 \pm 0.013$ & $0.987 \pm 0.013$ \cr
& $\ln{(10^{10} A_{s})}$ & $3.195 \pm 0.034$ & $3.186 \pm 0.035$ & $3.184 \pm 0.035$ & $3.170 \pm 0.037$ & $3.180 \pm 0.036$ & $3.169 \pm 0.037$ \cr
\hline
Extended & $N_{\rm eff}$ & --- & $3.87 \pm 0.42$ & --- & $4.00 \pm 0.43$ & $3.70 \pm 0.54$ & $3.99 \pm 0.59$ \cr
& $\sum{m_{\nu}}~\rm{[eV]}$ & --- & --- & $<0.45$ & $<0.67$ & --- & $<0.73$ \cr
& $Y_p$ & --- & --- & --- & --- & $0.277 \pm 0.037$ & $0.261 \pm 0.039$ \cr
\hline
Derived & $\sigma_8$ & $0.811 \pm 0.018$ & $0.862 \pm 0.033$ & $0.758 \pm 0.042$ & $0.798 \pm 0.053$ & $0.860 \pm 0.034$ & $0.796 \pm 0.055$ \cr
\hline
\hline
\end{tabular}
\begin{tablenotes}
\item Mean of the posterior distribution of cosmological parameters along with the symmetric 68\% confidence interval about the mean. We report the 95\% upper limit on the sum of neutrino masses $\sum{m_{\nu}}$. The primordial helium mass fraction $Y_p$ is enforced consistent with standard BBN unless we allow it to vary as a free parameter.
\end{tablenotes}
\label{table:lcdm}
\end{center}
\end{table*}

The neutrino sector is another area that the standard model is yet unable to fully describe, with open questions related to the effective number of neutrinos $\neff$ and their masses $m_{\nu}$.
The effective number of neutrinos is sensitive to both the number of neutrinos along with additional particle species that were relativistic at the photon decoupling epoch (e.g.~\cite{Komatsu,Mangano:2005cc}).
A joint analysis of CMB data from WMAP7 with baryon acoustic oscillation (BAO) distances from SDSS+2dF and Hubble constant from HST reveals a weak preference for extra relativistic species ($\neff = 4.34 \pm 0.87$)~\cite{Komatsu:2010fb}. When further combined with small-scale CMB data from ACT or SPT, this preference mildly increases and reaches the $2\sigma$ level ($\neff = 4.56 \pm 0.75$ with addition of ACT~\cite{dunkleyact} and $\neff = 3.86 \pm 0.42$ with addition of SPT~\cite{Keisler:2011aw}; further see~\cite{Hamann:2010bk,Archidiacono:2011gq,Riess:2011yx,Smith:2011es,Hou:2011ec, Hamann:2007pi, Hamann:2011hu, GonzalezMorales:2011ty, Calabrese:2011hg,Smith:2011ab, Hamann:2011ge,Giusarma:2011ex,Giusarma:2011zq,Fischler:2010xz,deHolanda:2010am,Nakayama:2010vs,Joudaki:2012uk}). 

A primary objective of this manuscript is to clarify how robust these recent
indications of additional light degrees of freedom are to assumptions of the underlying cosmology, in particular to alternative models of the dark energy, curvature of the universe, running of the spectral index, primordial helium abundance, and to the sum of neutrino masses, which we know is nonzero from neutrino oscillation experiments~\cite{hirata92, davis68, Eguchi:2002dm, Ahn:2002up}.

Constraining $\neff$ with cosmology is mainly achieved through tight CMB measurements of the 
redshift at matter-radiation equality~$z_{\rm{eq}}$, the baryon density $\Omega_b h^2$,
the angular size of the sound horizon~$\theta_s$, and the angular scale of photon diffusion~$\theta_d$~\cite{Hou:2011ec}. 
Keeping $z_{\rm{eq}}$ and $\Omega_b h^2$ fixed as $\neff$ increases
can be achieved by increasing the dark matter density $\Omega_c h^2$ (assuming massless neutrinos), which manifests in a large correlation with $\neff$ (shown in Fig.~\ref{fig:nuw}).
Meanwhile, an increase in $\neff$ and $Y_p$ both yield an enhanced Silk damping effect~\cite{Hu:1996vq,Hu:1995fqa,Bashinsky:2003tk,Hou:2011ec}, and by 
fixing $\theta_s$ it can be shown that $\theta_d \propto (1+f_\nu)^{0.22}/\sqrt{1+Y_p}$~\cite{Hou:2011ec}, where $f_\nu \equiv \rho_\nu/\rho_\gamma$ is proportional to $\neff$.
As a consequence, the suppression of the CMB damping tail can be picked out as a signature of extra relativistic species 
when $Y_p$ is known, while the constraints on $\neff$ are relaxed when allowing for $Y_p$ as a free parameter.

%A larger primordial helium abundance reduces the number density of free electrons at decoupling as helium recombines earlier than hydrogen, resulting in an increased mean free path of photons and thereby less power in the CMB damping tail~\cite{Hu:1995fqa,Komatsu:2010fb}. Some ways to achieve deviations in $Y_p$ include modifications to the expansion rate during BBN, energy injections due to annihilation or decay of heavy particles, and particle catalysis of BBN reactions~\cite{Jedamzik:2009uy}. 
	
An increase in $\neff$ further shifts the acoustic peak locations~\cite{Bashinsky:2003tk}, but this has been shown to be a small effect~\cite{Hou:2011ec}.
Instead, the constraint on $\neff$ can be improved by the inclusion of low-redshift distances and a prior on the Hubble constant, $H_0$, as these are useful in constraining $\Omega_c h^2$ and by extension $\neff$.
However, when allowing for evolving dark energy, the ability to improve constraints on $\neff$ from observations of the expansion history becomes diminished, as illustrated by the error ellipses for 
$\left\{\neff, \Omega_c h^2, w\right\}$ in Fig.~\ref{fig:nuw}. 
Therefore, the inclusion of SN data becomes critical to a precise determination of the effective number of neutrinos.

\begin{table*}[th]\footnotesize
\begin{center}
\caption{Constraints on Cosmological Parameters using SPT+WMAP+$H_0$+BAO.}
\begin{tabular}{lcccccccccccccc|}
\hline \hline
 & & $w$CDM & $\Lambda$CDM & $w$CDM & $\Lambda$CDM & $w$CDM & $w$CDM\\
 & & & +\neff+$\sum{m_\nu}$ & +\neff+$\sum{m_\nu}$ & +\neff+$\sum{m_\nu}$ & +\neff+$\sum{m_\nu}$ & +\neff+$\sum{m_\nu}$+$Y_p$\\
& & & & & +${dn_s \over d\ln k}+\Omega_k$ & +${dn_s \over d\ln k}+\Omega_k$ & +${dn_s \over d\ln k}+\Omega_k$\\
\hline \hline
Primary & $100 \Omega_b h^2	$ & $2.219 \pm 0.042$ & $2.272 \pm 0.043$ & $2.224 \pm 0.061$ & $2.244 \pm 0.054$ & $2.192 \pm 0.068$ & $2.168 \pm 0.079$ \cr
& $100\Omega_c h^2$ & $11.44 \pm 0.45$ & $13.05 \pm 0.94$ & $12.96 \pm 0.96$ & $13.2 \pm 1.0$ & $12.4 \pm 1.2$ & $13.1 \pm 1.7$ \cr
& $10^4\theta_s$ & $104.09 \pm 0.16$ & $103.95 \pm 0.18$ & $103.97 \pm 0.18$ & $103.97 \pm 0.19$ & $104.06 \pm 0.21$ & $103.83 \pm 0.44$ \cr
& $\tau$ & $0.083 \pm 0.014$ & $0.090 \pm 0.015$ & $0.086 \pm 0.015$ & $0.090 \pm 0.015$ & $0.088 \pm 0.015$ & $0.091 \pm 0.016$ \cr
& $n_s$ & $0.958 \pm 0.011$ & $0.987 \pm 0.013$ & $0.968 \pm 0.022$ & $0.978 \pm 0.015$ & $0.955 \pm 0.025$ & $0.949 \pm 0.027$ \cr
& $\ln{(10^{10} A_{s})}$ & $3.216 \pm 0.042$ & $3.170 \pm 0.037$ & $3.211 \pm 0.052$ & $3.179 \pm 0.045$ & $3.210 \pm 0.052$ & $3.200 \pm 0.057$ \cr
\hline
Extended & $w$ & $-1.10 \pm 0.11$ & --- & $-1.31 \pm 0.30$ & --- & $-1.46 \pm 0.39$ & $-1.35 \pm 0.41$ \cr
& $N_{\rm eff}$ & --- & $4.00 \pm 0.43$ & $3.59 \pm 0.57$ & $3.74 \pm 0.58$ & $3.10 \pm 0.74$ & $3.38 \pm 0.86$ \cr
& $\sum{m_{\nu}}~\rm{[eV]}$ & --- & $<0.67$ & $<1.2$ & $<1.2$ & $<1.2$ &  $<1.4$ \cr
& ${dn_s \over d\ln k}$ & --- & --- & --- & $-0.011 \pm 0.019$ & $-0.018 \pm 0.019$ & $-0.033 \pm 0.031$ \cr
& $100\Omega_k$ & --- & --- & --- & $0.75 \pm 0.93$ & $0.13 \pm 0.99$ & $0.76 \pm 1.5$ \cr
& $Y_p$ & --- & --- & --- & --- & --- & $0.196 \pm 0.084$ \cr
\hline
Derived & $\sigma_8$ & $0.848 \pm 0.049$ & $0.798 \pm 0.053$ & $0.775 \pm 0.063$ & $0.768 \pm 0.070$ & $0.803 \pm 0.085$ &  $0.779 \pm 0.091$ \cr
\hline
\hline
\end{tabular}
\begin{tablenotes}
\item 
Same as Table~\ref{table:lcdm} but with the addition of $\{w,\nrun,\Omega_k\}$. 
Due to the large correlation between $n_s$ and $\nrun$ at our pivot scale $k_0 = 0.002/{\rm{Mpc}}$, we quote values for $n_s$ at a less correlated scale $k_0 = 0.015/{\rm{Mpc}}$.
For the ``$w$CDM$+\neff+\sumnu+\nrun+\Omega_k$" case where $\neff$ is closest to to the boundary at 3, we also considered a run where we impose a hard prior of $\neff \geq 3$. Here, we find $\neff = {{3.65}~^{3.82,~4.62}_{3.00,~3.00}}$, where the two sets of upper and lower boundaries denote 68\% and 95\% CLs, respectively. The changes to the sum of neutrino masses and other parameters that weakly correlate with $\neff$ are small ($<10\%$). 
While all within 1$\sigma$, the largest changes are seen in $100\Omega_c h^2 = 13.17 \pm 0.97$ (compared to $100\Omega_c h^2 = 12.4 \pm 1.2$), $w = -1.25 \pm 0.30$ (compared to $w = -1.46 \pm 0.39$), $n_s = 0.971 \pm 0.019$ (compared to $n_s = 0.955 \pm 0.025$), $\nrun = -0.0088 \pm 0.0168$ (compared to $\nrun = -0.018 \pm 0.019$), and $100\Omega_k = 0.2 \pm 1.1$ (compared to $100\Omega_k = 0.13 \pm 0.99$). 
This particular configuration of parameter space and datasets shows the largest extent to which parameters may change with an $\neff>3$ prior as compared to our other runs. 
The changes to the parameters are more modest when including SNe because of the preference for larger values of $\neff$, as seen in Table~\ref{table:wcdmsn}.
\end{tablenotes}
\label{table:wcdm}
\end{center}
\vspace{-0.2em}
\end{table*}

The dark energy equation of state (EOS) is moreover anti-correlated with the sum of neutrino masses~\cite{Hannestad:2005gj,Komatsu,Ichikawa:2004zi,Komatsu:2010fb}.
In the CMB temperature power spectrum, the sum of neutrino masses shifts the first peak position to lower multipoles by changing the fraction of matter to radiation at decoupling, which can be compensated by a reduction in the Hubble constant (similar to the case for positive universal curvature)~\cite{Hannestad:2003xv,Ichikawa:2004zi,Komatsu}. 
BAO distances and an $H_0$ prior can therefore be used to reduce correlations between the sum of neutrino masses and the dark energy EOS, but also the curvature density.

The strongest limits on the sum of neutrino masses from the CMB combined with probes of the expansion history and matter power spectrum place it at sub-eV level~\cite{Hannestad:2003xv,Seljak:2006bg, Goobar:2006xz, Ichiki:2008ye, Tereno:2008mm,  Komatsu, Komatsu:2010fb, Reid:2009xm, Thomas:2009ae, dePutter:2012sh,Spergel:2003cb,Spergel:2006hy,Benson:2011ut,RiemerSorensen:2011fe,Reid:2009nq,Tegmark:2003ud,Crotty:2004gm,Barger:2003vs,Allen:2003pta,Elgaroy:2004rc,Hannestad:2006mi,Xia:2012na}. 
We take the conservative approach in only combining CMB data with low-redshift measurements of the expansion history.
While SN observations play an important role in constraining the dark energy EOS and thereby reduce the correlation between $\sumnu$ and $w$, these observations are not powerful in constraining the curvature of the universe and therefore less helpful in reducing the correlation between $\sumnu$ and $\Omega_k$ (e.g.~\cite{RiessSNe, Perlmutter,Komatsu}).

Beyond the vanilla parameters and the three additional parameters $\{\neff, \sumnu, w\}$, we relax the commonly employed strong inflation prior on the universal curvature $\Omega_k$ and running of the spectral index $\nrun$. 
Given that most popular models of inflation predict $|{\nrun}| \lsim 10^{-3}$~\cite{Kosowsky:1995aa,Baumann:2008aq} and $|\Omega_k| \lsim 10^{-4}$ (e.g.~\cite{liddlelyth,weinberg,Baumann:2008aq}), at the level of precision of present CMB data it is generally justified to fix these two parameters to their fiducial values of zero. 
However, given the mild preference for $\neff > 3$~\cite{Komatsu:2010fb, dunkleyact, Keisler:2011aw}, we allow for the possible existence of inflationary models with large curvature or running.
In particular, $|\Omega_k| \sim 10^{-2}$ may be generated in models of open inflation in the context of string cosmology~\cite{Freivogel:2005vv,Baumann:2008aq}, while a large negative running may be produced by multiple fields, temporary breakdown of slow-roll, or several distinct stages of inflation~\cite{Easther:2006tv,Burgess:2005sb,Joy:2007na,Baumann:2008aq}.

Among alternatives to the cosmological constant with $w = -1$, the most popular are scalar field models with potentials tailored to give rise to late-time acceleration and current equation of state for the dark energy, $w$, close to -1~\cite{Ford:1987de, Ratra:1987rm, Wetterich88, Peebles:1987ek, Zlatev, Ferreira, Caldwell:1997ii, Chiba}.
Like a cosmological constant, these models are fine-tuned to have dark energy dominate today. 
However, the requirement $w \gtrsim -1$ currently, does not imply that dark energy was negligible at
earlier times, specifically redshift $z \gtrsim 2$, where we have no direct constraints. 
Given the degeneracy between dark energy and the sum of neutrino masses,
we further consider a model that describes dark energy as non-negligible in the early universe in Sec.~\ref{subformafive}.

We describe our analysis method in Section~\ref{methform}. In Section~\ref{resultlabel}, we provide constraints on a $\Lambda$CDM model with three massive neutrinos and additional light degrees of freedom, then follow up with successive additions of a constant dark energy equation of state, universal curvature, running of the spectral index, and primordial helium abundance (all parameters defined in Table~\ref{table:priors}).
We also explore the constraints for a time-varying dark energy equation of state, including an early dark energy model. Lastly, we compare the constraints from present data to the constraints expected from the Planck experiment.
Section~\ref{conclusionlabel} concludes with a discussion of our findings. 

%%%%%%%%%%%%%%%%%%%%%%%%%%%%%
\begin{table*}[th]\footnotesize
\begin{center}
\caption{Constraints on Cosmological Parameters using SPT+WMAP+$H_0$+BAO+SNe.}
\begin{tabular}{lcccccccccccccc|}
\hline \hline
 & & $w$CDM & $w$CDM & $w$CDM & $w$CDM\\
 & & & +\neff+$\sum{m_\nu}$ & +\neff+$\sum{m_\nu}$ & +\neff+$\sum{m_\nu}$+$Y_p$\\
& & & & +${dn_s \over d\ln k}+\Omega_k$ & +${dn_s \over d\ln k}+\Omega_k$\\
\hline \hline
Primary & $100 \Omega_b h^2$ & $2.223 \pm 0.041$ & $2.257 \pm 0.048$ & $2.226 \pm 0.059$ & $2.171 \pm 0.080$ \cr
& $100\Omega_c h^2$ & $11.36 \pm 0.41$ & $13.14 \pm 0.94$ & $13.1 \pm 1.0$ & $14.0 \pm 1.4$   \cr
& $10^4\theta_s$ & $104.10 \pm 0.16$ & $103.95 \pm 0.17$ &  $103.99 \pm 0.18$ & $103.66 \pm 0.36$ \cr
& $\tau$ & $0.082 \pm 0.014$ & $0.088 \pm 0.015$ & $0.089 \pm 0.016$ & $0.090 \pm 0.015$ \cr
& $n_s$ & $0.960 \pm 0.010$ & $0.981 \pm 0.015$ & $0.970 \pm 0.019$ & $0.953 \pm 0.026$ \cr
& $\ln{(10^{10} A_{s})}$ & $3.209 \pm 0.039$ & $3.185 \pm 0.041$ & $3.194 \pm 0.049$ & $3.197 \pm 0.053$ \cr
\hline
Extended & $w$ & $-1.049 \pm 0.072$ & $-1.09 \pm 0.11$ & $-1.10 \pm 0.11$ & $-1.13 \pm 0.12$ \cr
& $N_{\rm eff}$ & --- & $3.88 \pm 0.44$ & $3.58 \pm 0.60$ & $3.78 \pm 0.61$ \cr
& $\sum{m_{\nu}}~\rm{[eV]}$ & --- & $<0.92$ & $<1.2$ & $<1.7$ \cr
& ${dn_s \over d\ln k}$ & --- & --- & $-0.013 \pm 0.019$ & $-0.035 \pm 0.030$ \cr
& $100\Omega_k$ & --- & --- & $0.64 \pm 0.95$ & $1.2 \pm 1.1$ \cr
& $Y_p$ & --- & --- & --- & $0.176 \pm 0.079$ \cr
\hline
Derived & $\sigma_8$ & $0.830 \pm 0.038$ &  $0.790 \pm 0.060$ & $0.774 \pm 0.072$ & $0.751 \pm 0.081$ \cr
\hline
\hline
\end{tabular}
\begin{tablenotes}
\item 
Same as Table~\ref{table:wcdm} but with the addition of supernova distance measurements from the Union2 compilation.
For the case ``$w$CDM$+\neff$+$\sumnu$$+\nrun$$+\Omega_k$," we also considered a run where we impose a hard prior of $\neff \geq 3$. Here, we find $\neff = {{3.74}~^{3.92,~4.68}_{3.00,~3.00}}$, where the two sets of upper and lower boundaries denote 68\% and 95\% CLs, respectively. The largest changes this prior induces in other parameters are in $n_s = 0.974 \pm 0.017$ (compared to $n_s = 0.970 \pm 0.019$), $\nrun = -0.010 \pm 0.017$ (compared to $\nrun = -0.013 \pm 0.019$), and $100\Omega_k = 0.52 \pm 0.92$ (compared to $100\Omega_k = 0.64 \pm 0.95$). All other parameters are modestly affected by our choice of prior ($<10\%$).
\end{tablenotes}
\label{table:wcdmsn}
\end{center}
\end{table*}
%%%%%%%%%%%%%%%%%%%%%%%%%%%%%

\section{Methodology}
\label{methform}

We employed a modified version of CosmoMC~\cite{Lewis:2002ah,cosmomclink} in performing Markov Chain Monte Carlo (MCMC) analyses of extended parameter spaces with CMB data from WMAP7~\cite{Komatsu:2010fb} and SPT~\cite{Keisler:2011aw}, BAO distance measurements from SDSS+2dFGRS~\cite{Percival:2009xn}, the Hubble constant from HST~\cite{Riess:2011yx}, and SN distances from the SCP Union2 compilation~\cite{Amanullah:2010vv}. 
In determining the convergence of our chains, we used the Gelman and Rubin $R$ statistic~\cite{gelmanrubin}, where $R$ is defined as the variance of chain means divided by the mean of chain variances. 
To stop the runs, we generally required the conservative limit $(R - 1) < 10^{-2}$, 
and checked that further exploration of the tails does not change our results.

The CMB temperature and E-mode polarization power spectra were obtained from a modified version of the Boltzmann code CAMB~\cite{LCL,camblink}. 
We approximated the effect of a dark energy component with a time-varying EOS by incorporating the PPF module by Fang, Hu, \& Lewis~(2008)~\cite{Fang} into CosmoMC.
Given that the small scale CMB measurements of SPT come with much smaller error bars than ACT~\cite{dunkleyact,Keisler:2011aw}, the further 
inclusion of the ACT dataset would not lead to significant improvements in our constraints, as we explicitly checked.

All parameters are defined in Table~\ref{table:priors}. In our analyses, we always include the ``vanilla" parameters, given by the full set $\left\{{\Omega_{b}h^2, \Omega_{c}h^2, \theta_s, \tau, n_{s}, \ln{(10^{10} A_{s})}}\right\}$.
Our constraints correspond to the mean of the posterior distribution of cosmological parameters along with the symmetric 68\% confidence interval about the mean. 
We impose uniform priors on the cosmological parameters, and let the prior ranges to be significantly larger than the posterior, such that the parameter estimates are unaffected by the priors.
When including the sum of neutrino masses and early dark energy density, we report the 95\% upper limit on these parameters.

When allowing for nonzero neutrino rest mass, we distribute the sum of neutrino masses ($\sumnu = 94~{\rm{eV}}~\Omega_{\nu} h^2$) equally among 3 active neutrinos.
We treat additional contributions to $\neff$ as massless, such that $\neff = (3 + N_{\rm{ml}})$, where $N_{\rm{ml}}$ denotes the massless degrees of freedom. In principle, these additional states could be massive, for example see recent treatments in Refs.~\cite{Hamann:2010bk, Hamann:2011ge,Joudaki:2012uk}. 

\begin{figure*}[!t]
\begin{center}
\vspace{-0.3em}
\includegraphics[bb=129.851996 231.434063 456.983916 539.338976,clip,scale=0.5]{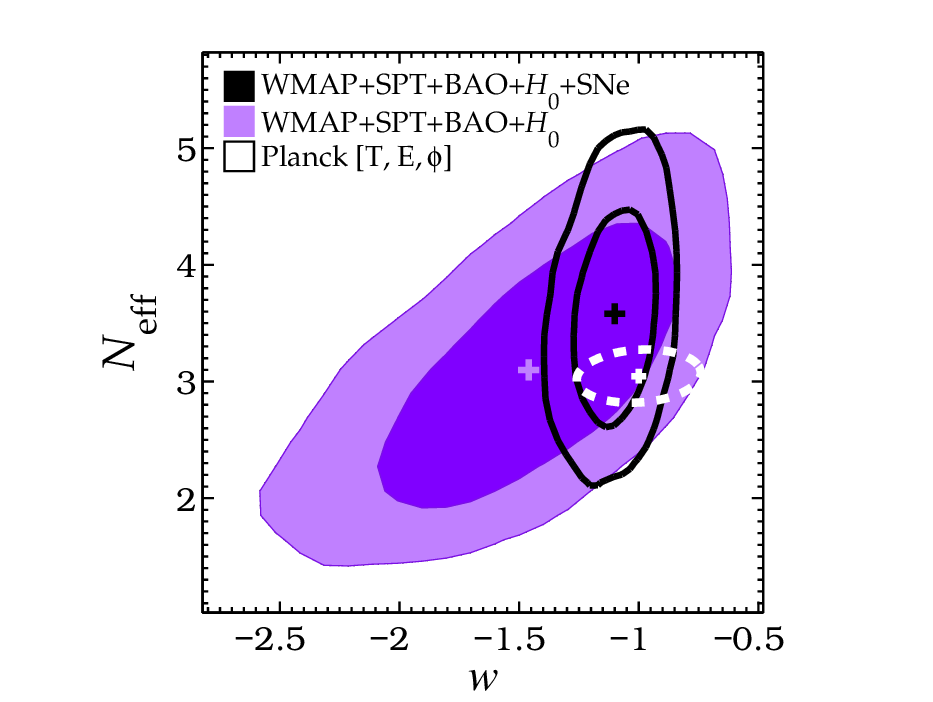}
\includegraphics[bb=181.075002 230.587501 444.496979 539.338976,clip,scale=0.5]{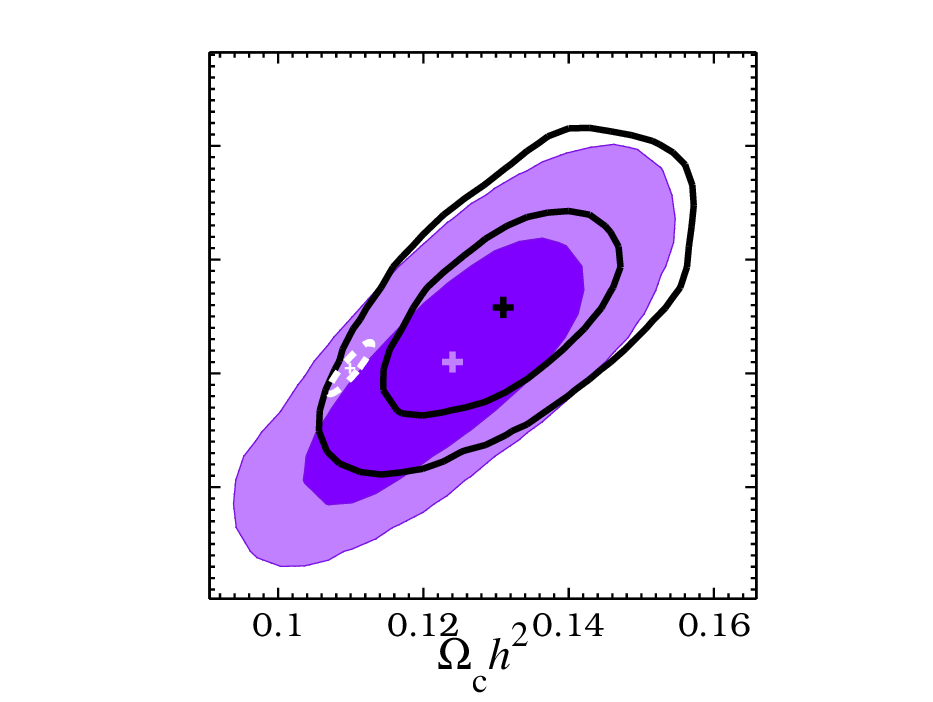}
\includegraphics[bb=180.499002 230.534344 445.090979 539.338976,clip,scale=0.5]{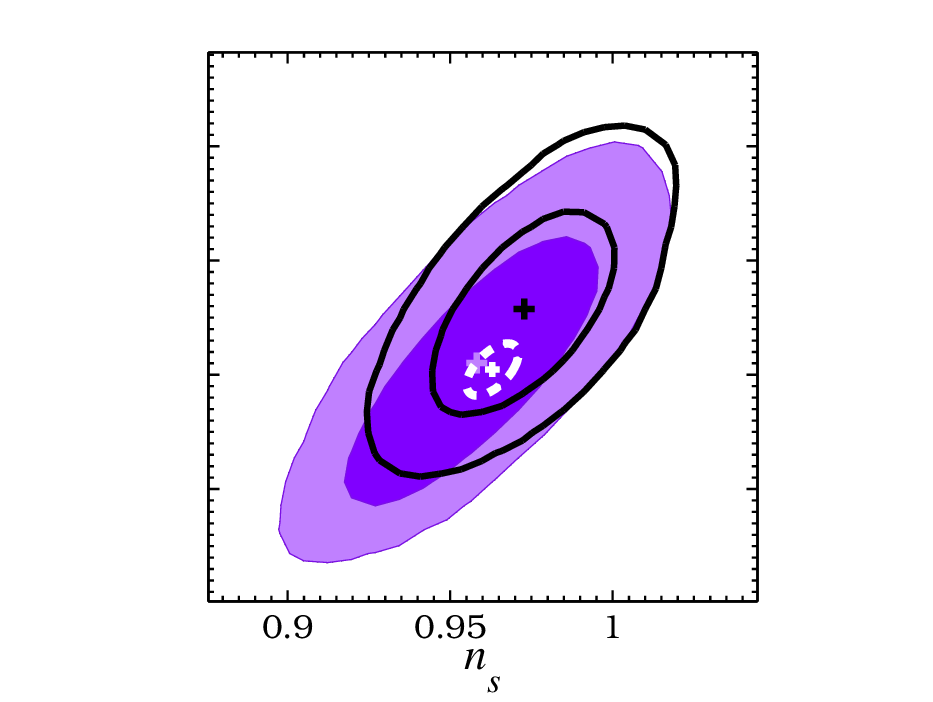}
\includegraphics[bb=113.952161 230.434063 456.983916 539.410976,clip,scale=0.5]{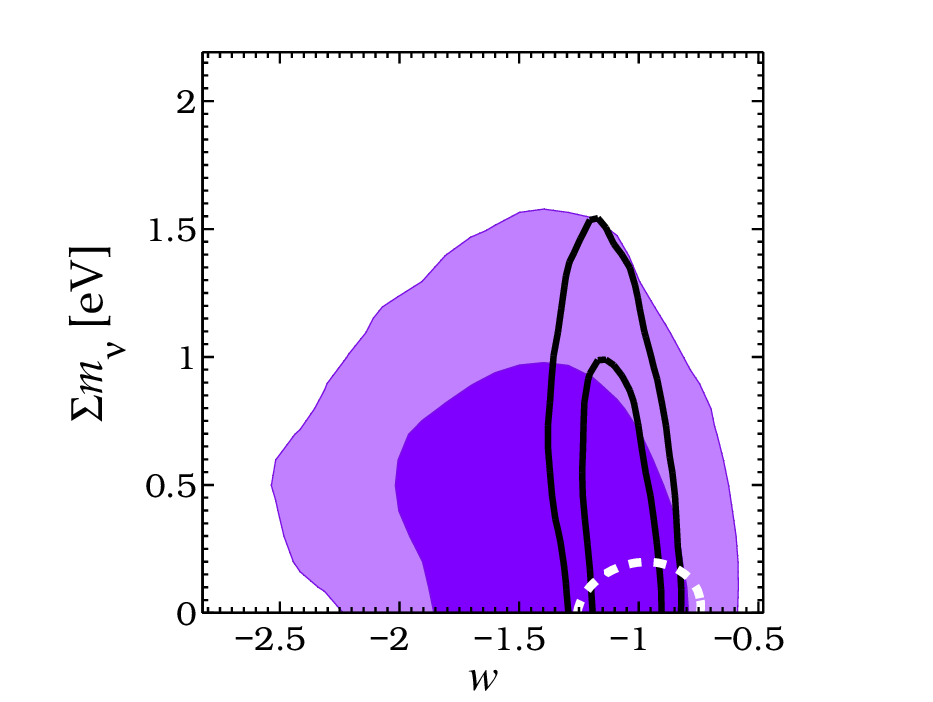}
\includegraphics[bb=181.075002 229.587501 444.496979 539.410976,clip,scale=0.5]{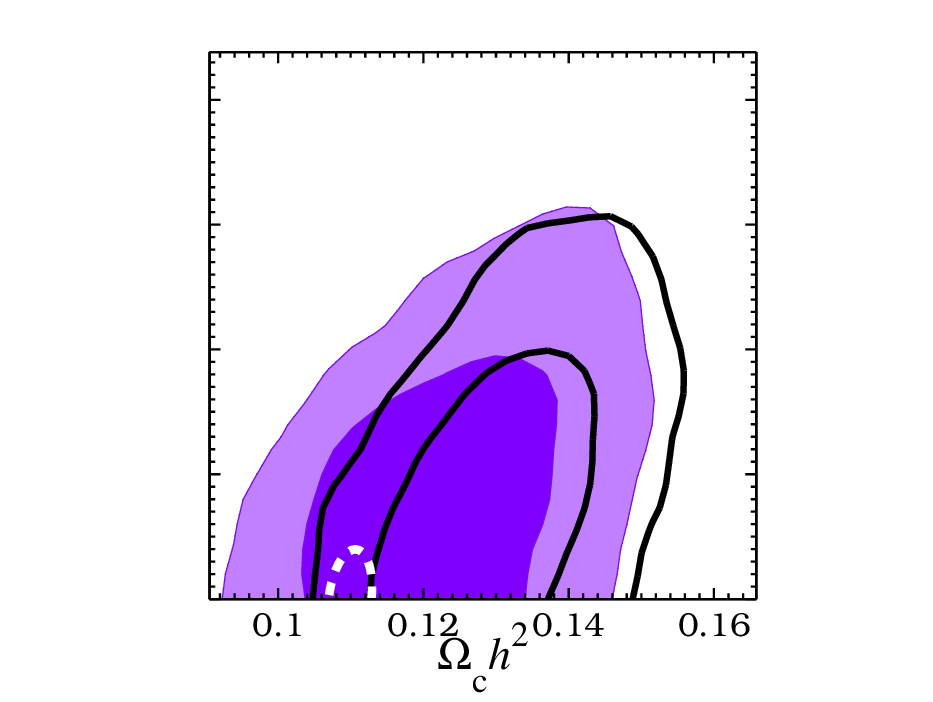}
\includegraphics[bb=180.661002 229.587501 445.000979 539.410976,clip,scale=0.5]{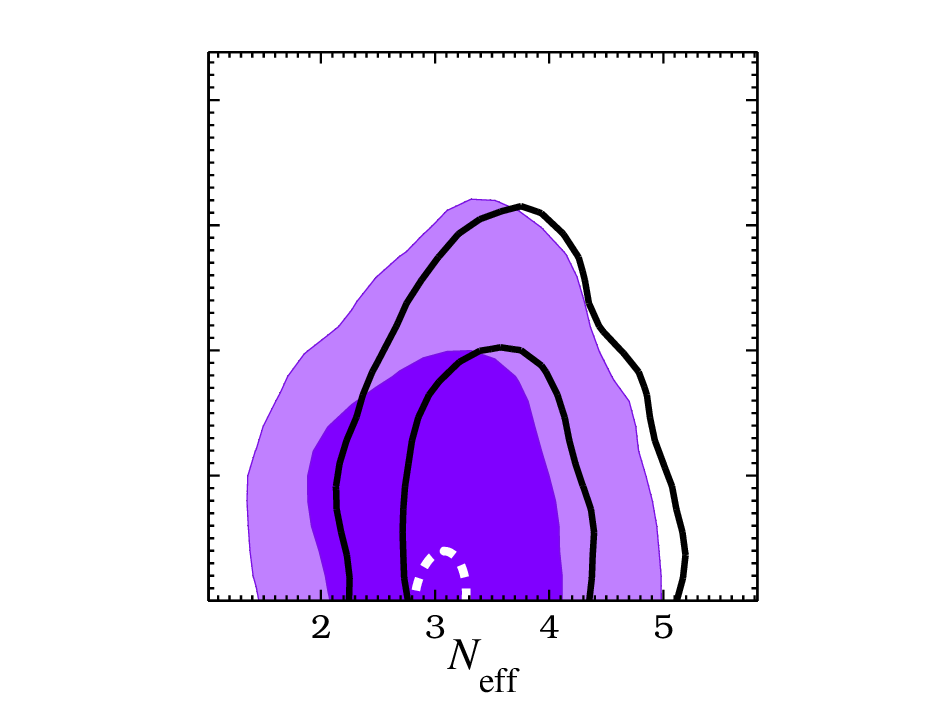}
\end{center}
\vspace{-2.1em}
\caption{Joint two-dimensional marginalized constraints (inner 68\% CL, outer 95\% CL) for the extended parameter combination ``vanilla$+\neff$$+\sum{m_{\nu}}$$+w$$+\Omega_k$$+\nrun$," showing $\neff$ against $\{w, \Omega_c h^2, n_s\}$ and $\sumnu$ against $\{w, \Omega_c h^2, \neff\}$. The purple shaded confidence regions are for ``WMAP+SPT+$H_0$+BAO" and the regions enclosed by black lines are for ``WMAP+SPT+$H_0$+BAO+SNe," where the BAOs and SNe are from SDSS+2dF and Union2, respectively. For Planck (T, E, $\phi$) in dashed white, we have centered the $1\sigma$ error ellipses on the fiducial values of the parameters that went into computing the Fisher matrix. The exception to this convention is $\sumnu$, which we have shifted down to 0 eV from its fiducial value of 0.17 eV for simpler visual comparison with the upper bounds from present data.}
\label{fig:nuw}
\end{figure*}

\begin{figure*}[!t]
\begin{center}
\vspace{-0.3em}
\includegraphics[bb=98.496349 230.587501 445.000979 539.410976,clip,scale=0.5]{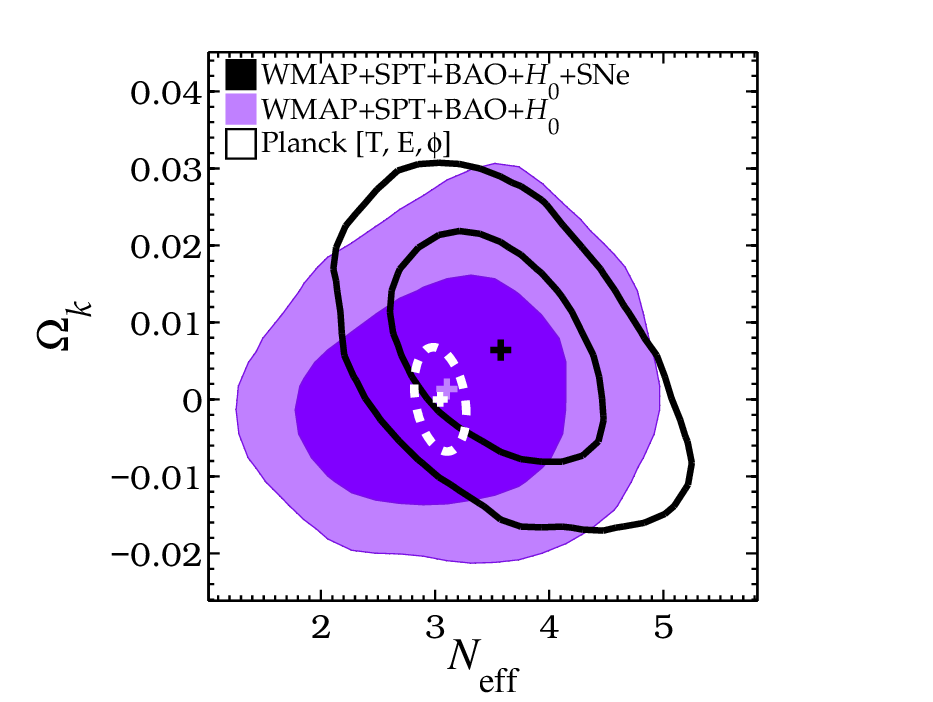}
\includegraphics[bb=177.228065 231.776344 444.586979 539.338976,clip,scale=0.5]{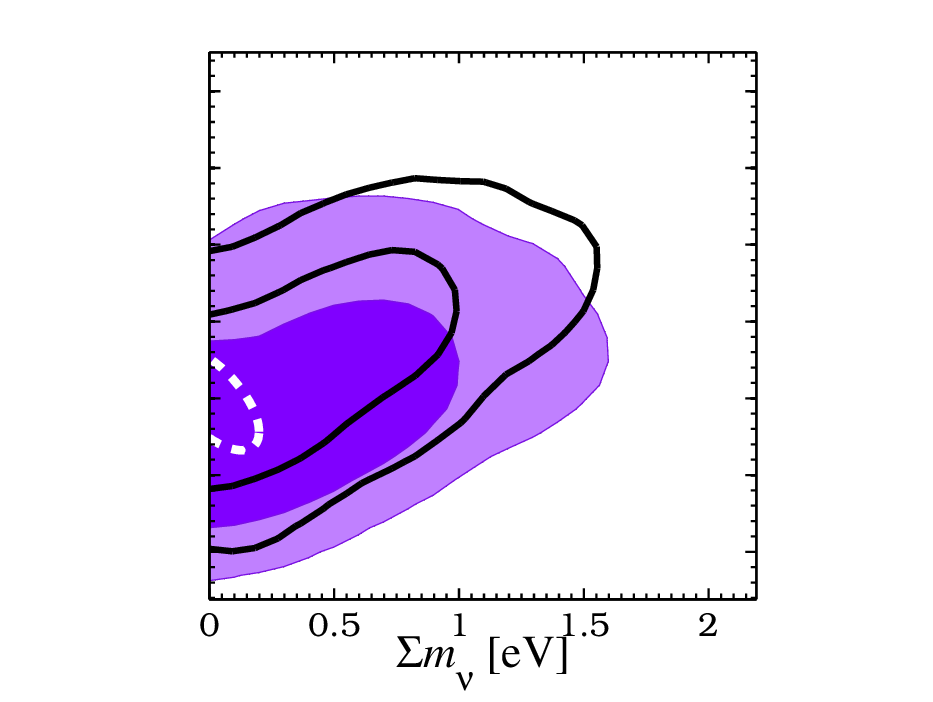}
\includegraphics[bb=180.409002 230.534344 459.053916 539.410976,clip,scale=0.5]{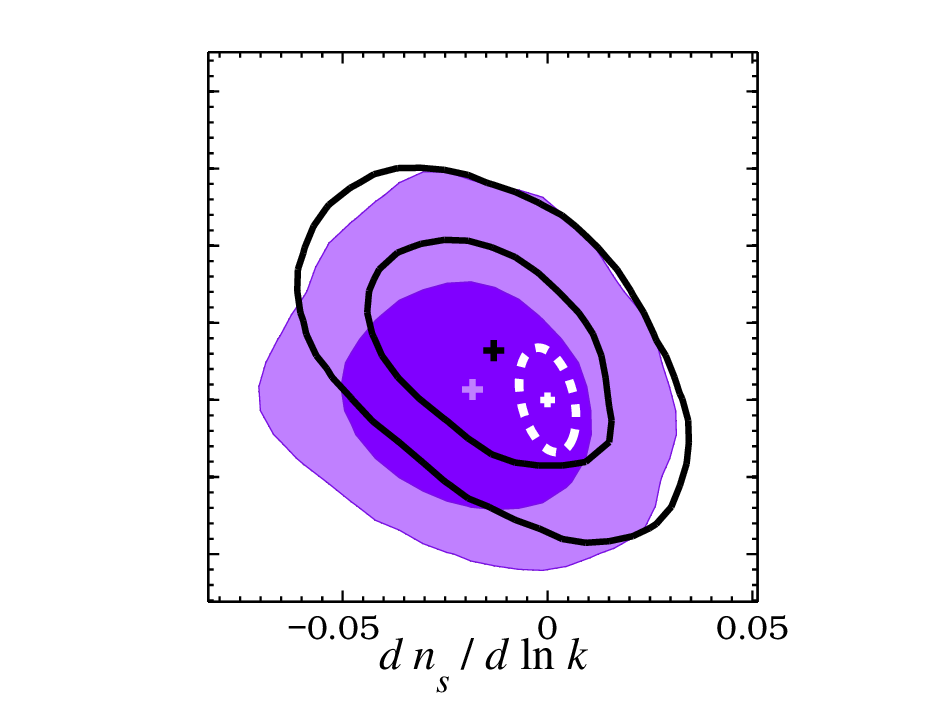}
\includegraphics[bb=97.454106 229.587501 445.000979 539.338976,clip,scale=0.5]{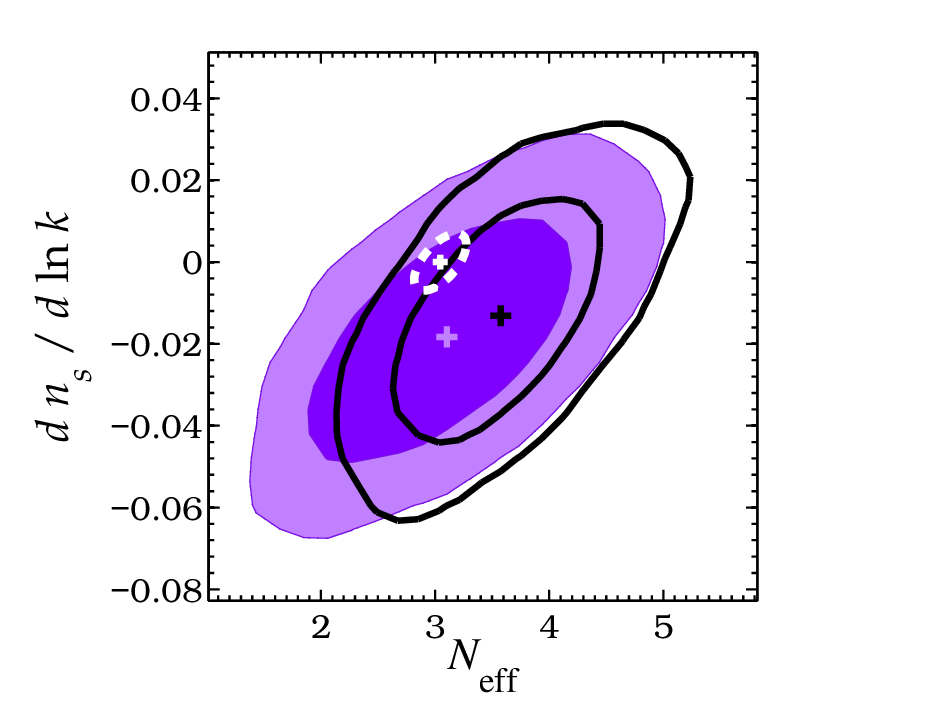}
\includegraphics[bb=177.228065 230.776344 444.586979 539.410976,clip,scale=0.5]{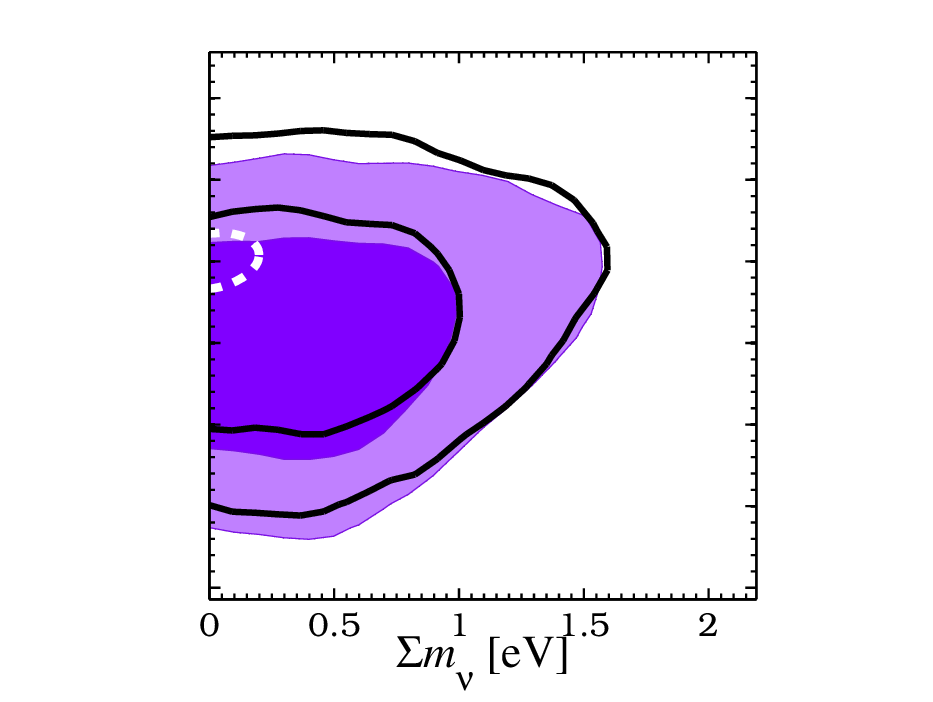}
\includegraphics[bb=177.673002 226.204345 448.168979 541.912976,clip,scale=0.493]{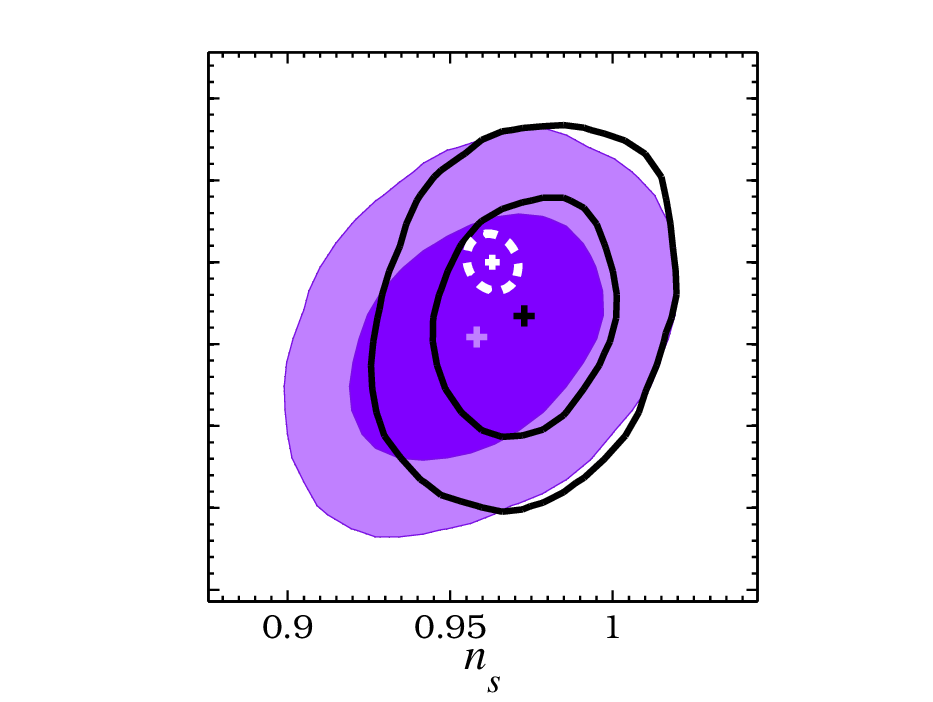}
\end{center}
\vspace{-2.1em}
\caption{Same as Fig.~\ref{fig:nuw}, but for $\Omega_k$ against $\{\neff, \sumnu, \nrun\}$ and $\nrun$ against $\{\neff, \sumnu, n_s\}$.
}
\label{fig:omkrun}
\end{figure*}

Since we impose $1.047<\neff<10$, the number of relativistic species is always positive at early times. 
At late times, our prior on $\neff$ implies that the number of relativistic species can be negative when the three active neutrinos are massive ($-1.953 < N_{\rm{ml}} < 7$). However, the total radiation energy density is always positive (at late times ${\propto 1 + 0.227 N_{\rm{ml}}}$ when the three active neutrinos are massive). We choose this particular prior on $\neff$ in order for the data itself to rule out a given part of parameter space.
In Figs~\ref{fig:nuw}-\ref{fig:ede}, 
we find that the marginalized contours on $\neff$ close before the lower end of our prior, such that the data itself is constraining the radiation content from below. For completeness, we also considered several conventional runs with the prior $\neff \geq 3$, such that $N_{\rm{ml}} \geq 0$, and we find no qualitative changes in our results. For complete details, see the captions of Tables~\ref{table:wcdm},~\ref{table:wcdmsn},~\ref{table:edecdmsn}.

As part of our analysis of extended parameter spaces, we consider cases with the primordial fraction of baryonic mass in helium $Y_p$ as an unknown parameter to be determined by the data.
However, when we do not allow $Y_p$ to vary freely, it is determined in a BBN-consistent manner within CAMB via the PArthENoPE code~\cite{Pisanti:2007hk}, 
which enforces
\begin{equation}
\yhe = 0.2485 + 0.0016\left[{(273.9\Omega_bh^2-6)+100(S-1)}\right].
\label{eqn:yp}
\end{equation}
Here $S = \sqrt{1 + (7/43)\Delta N_\nu}$ encapsulates deviations from standard BBN~\cite{Kneller:2004jz,Simha:2008zj,Steigman:2007xt}, 
and we let $\Delta N_\nu = (\neff - 3.046)$ in agreement with the SPT analysis. Aside from the derived limits on $Y_p$, we explicitly checked that our results do not significantly change ($<10\%$) when  
passing $\Delta N_\nu = 0$ to PArthENoPE instead.

Furthermore, in our analysis we either consider ``enforcing the strong inflation prior" on the curvature and running, by which $\{\Omega_k \equiv 0, \nrun \equiv 0\}$, or ``relaxing the strong inflation prior" such that $\{\Omega_k, \nrun\}$ are allowed to vary as free parameters to be constrained by the data.
We define the running of the spectral index via the dimensionless power spectrum of primordial curvature perturbations:
\begin{equation}
\Delta^{2}_{R}(k) = \Delta^{2}_{R}(k_0)\left(\frac{k}{k_0}\right)^{n_s-1+\frac{1}{2}\ln(k/k_0) dn_s/d\ln k},
\end{equation}
where the pivot scale $k_0 = 0.002/{\rm{Mpc}}$. Due to the large correlation between $n_s$ and $\nrun$ at this scale, we always quote our values for $n_s$ at a scale $k_0 = 0.015/{\rm{Mpc}}$, where the tilt and running are less correlated, such that 
$n_s(k_0=0.015/{\rm{Mpc}}$) = $n_s(k_0=0.002/{\rm{Mpc}}$$) + \ln(0.015/0.002)dn_s/d\ln k$~\cite{Cortes:2007ak}. An example of the remaining correlation between the spectral index and its running is shown in Fig.~\ref{fig:omkrun}.

\begin{figure}[!t]
%\vspace{-0.37cm}
\vspace{-0.3cm}
\epsfxsize=3.7in
\centerline{\epsfbox{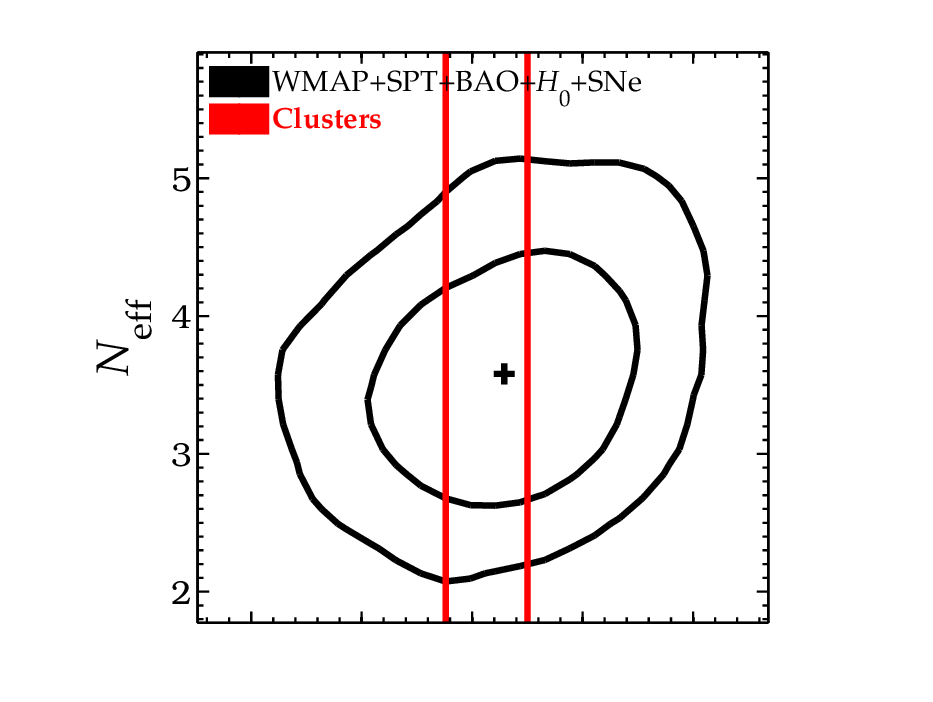}}
\vspace{-0.83cm}
\centerline{\epsfbox{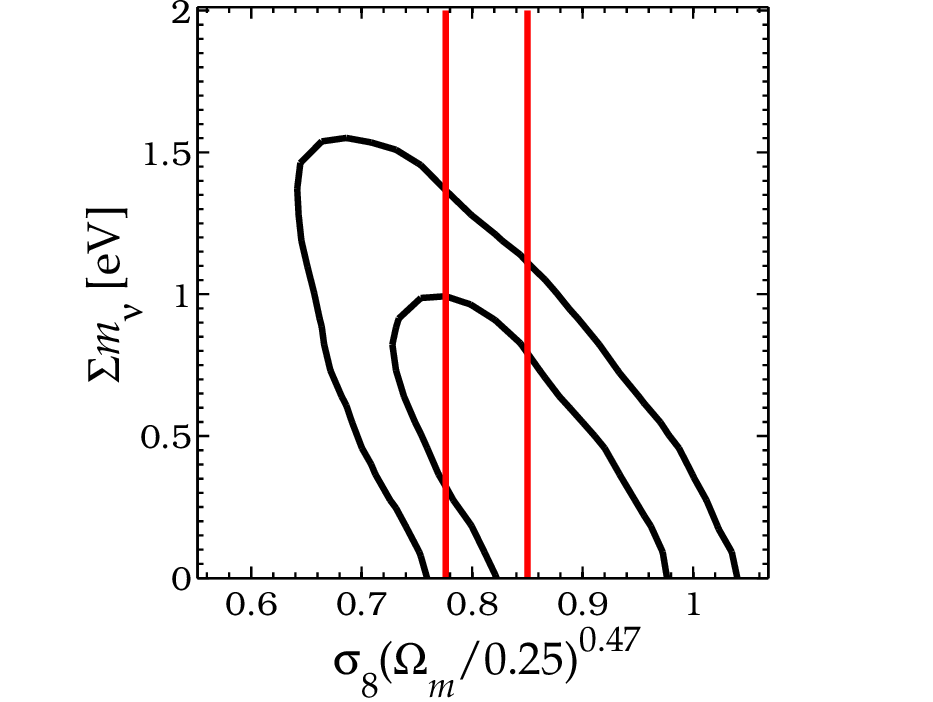}}
%\vspace{-0.25cm}
\vspace{-0.2cm}
\caption{Joint two-dimensional marginalized constraints on $\sigma_8 (\Omega_m/0.25)^{0.47}$ against $\{\neff, \sumnu\}$. The black confidence regions (inner 68\%, outer 95\%) are for the extended parameter combination ``vanilla$+\neff$$+\sum{m_{\nu}}$$+w$$+\Omega_k$$+\nrun$" using~the~data from ``WMAP7+SPT+$H_0$+BAO+SNe," while the vertical red lines denote 
the 68\% confidence interval about the mean from the local ($0.025 < z < 0.25$) galaxy cluster abundance measurement of Vikhlinin~et~al.~(2009)~\cite{Vikhlinin:2008ym}.}
\label{fig:sigma8}
\end{figure}
 
\section{Results}
\label{resultlabel}

We now explore the constraints on extended parameter spaces with the CMB (WMAP7+SPT), BAO distances (SDSS+2dFGRS), and an HST prior on the Hubble constant. Beginning with Sec.~\ref{subformathree} we always also consider SN distance measurements from the Union2 compilation. 
In Sec.~\ref{subformasix}, we discuss the expected constraints from Planck~\cite{planckbb, plancksite}.

\subsection{$\Lambda$CDM with Massive Neutrinos}
\label{subformaone}

\subsubsection{Enforcing the inflation prior on $\{\Omega_k,\nrun\}$}
In Table~\ref{table:lcdm}, we begin by allowing the effective number of neutrinos, sum of neutrino masses, and primordial helium abundance to vary as free parameters, both separately and jointly, in a $\Lambda$CDM universe. 

First for $\Lambda$CDM alone, then with $\neff$ and $Y_p$ added separately, we reproduce the results in Ref.~\cite{Keisler:2011aw}. 
In particular, with $\neff = 3.87 \pm 0.42$ in the parameter space given by ``vanilla$+\neff$," we recover the reported $2\sigma$ deviation from canonical $\neff = 3.046$~\cite{Keisler:2011aw, dunkleyact}. 
Given the well known degeneracy between $\neff$ and $Y_p$~\cite{Keisler:2011aw,dunkleyact,Hou:2011ec} (also see discussion in Sec.~\ref{introlabel}), we find $\neff = 3.70 \pm 0.54$ when further allowing $Y_p$ to vary as a free parameter irrespectively of the BBN expectation.
Allowing for three active neutrinos to have mass, and treating additional contributions to $\neff$ as massless,
we find an even larger deviation with the standard value as $\neff = 4.00 \pm 0.43$ for the parameter combination ``vanilla$+\neff$+$\sumnu$" (consistent with~\cite{Keisler:2011aw}). 

Here, the upper bound on the sum of neutrino masses is 0.67~eV (95\% CL) and we find the spectral index to be consistent with unity within 1$\sigma$ ($n_s = 0.987 \pm 0.013$). 
The neutrino mass constraint is to be compared with 0.45~eV at 95\% CL in ``vanilla+$\sumnu$" (consistent with~\cite{Riess:2011yx}). This 0.45 eV constraint is competitive with 
the robust upper bound of 0.36 eV when including CMASS~\cite{dePutter:2012sh}, the conservative upper bound of 0.34 eV from the MegaZ photometric redshift catalog of luminous red galaxies~\cite{Thomas:2009ae}, and the conservative upper bound of 0.41 eV from the CFHTLS galaxy angular power spectrum~\cite{Xia:2012na}.

When we consider ``vanilla$+\neff$+$\sumnu$$+Y_p$," the statistical significance of the $\neff$ deviation is reduced from $2.2\sigma$ to $1.6\sigma$, 
and the upper bound on the sum of neutrino masses moderately weakens to 0.73~eV (95\% CL). 
While the primordial helium abundance from the CMB+BAO+$H_0$ has been found mildly in tension $\left({\sim2\sigma}\right)$~\cite{dunkleyact,Keisler:2011aw} with that 
from observations of metal-poor extragalactic H II regions~\cite{Peimbert:2007vm,Izotov:2007ed,Izotov:2010ca,Aver:2010wq,Aver:2010wd,Aver:2011bw}, we find constraints on $Y_p$ consistent to within $1\sigma$ with these observations. This is mainly due to the strong negative correlation between $Y_p$ and $\neff$ (as reported in~\cite{dunkleyact,Keisler:2011aw,Hou:2011ec} and detailed in Sec.~\ref{introlabel}).
For instance, Aver, Olive, \& Skillman~(2011)~\cite{Aver:2011bw} determine $Y_p = 0.2534 \pm 0.0083$ via an MCMC analysis that accounts for both statistical and systematic uncertainties, which agrees with $Y_p = 0.277 \pm 0.037$ in ``vanilla$+\neff$$+Y_p$" and with $Y_p = 0.261 \pm 0.039$ in ``vanilla$+\neff$+$\sumnu$$+Y_p$."

Moreover, when $\neff$ and $\sumnu$ are analyzed in a joint setting, we find that the data is both consistent with higher values of $\Omega_c h^2 = 0.13 \pm 0.01$ and lower values of
$\sigma_8 = 0.80 \pm 0.05$, which perfectly agrees with low-redshift measurements of $\sigma_8$ from the abundance of clusters~\cite{Vikhlinin:2008ym, Rozo:2009jj,Vanderlinde:2010eb,Mantz:2009fw,Sehgal:2010ca} 
(as also noted in Ref.~\cite{Keisler:2011aw}). 
This is because the amount of suppression in matter clustering by the free-streaming of light neutrinos increases with mass~\cite{Bond:1980ha,Bond:1983hb,Ma:1996za,Spergel:2006hy}, which gives a large anti-correlation between $\sumnu$ and $\sigma_8$, an example of which can be seen in Fig.~\ref{fig:sigma8}.

\subsubsection{Relaxing the inflation prior on $\{\Omega_k,\nrun\}$}
\label{subformajaja}

Let us now relax the strong inflation prior on the curvature of the universe and running of the spectral index by considering the parameter combination ``vanilla$+\neff$+$\sumnu$$+\nrun$$+\Omega_k$" in Table~\ref{table:wcdm}.

Here, $\neff$ becomes increasingly consistent with the canonical value at $1.2\sigma$ (down from $2.2\sigma$), mainly as a result of the positive correlation with $\nrun$, which
also brings the tilt down to $n_s = 0.978 \pm 0.015$ (from $n_s = 0.987 \pm 0.013$).
Further, we find that the correlation between $\sumnu$ and $\Omega_k$ degrades the upper bound on the sum of neutrino masses by close to a factor of 2 to $\sumnu < 1.2~{\rm{eV}}$ (95\% CL).
As a consequence of the well known anti-correlation with the sum of neutrino masses,  which increases when relaxing the strong inflation prior, 
the amplitude of matter fluctuations is seen to prefer smaller values at $\sigma_8 = 0.768 \pm 0.070$ (as compared to $\sigma_8  = 0.798 \pm 0.053$ when $\nrun$ and $\Omega_k$ are held fixed).

However, as the strong inflation prior is relaxed, both the running and curvature are consistent with zero to $1\sigma$. It is therefore far from certain that shifts in parameters 
other than $\{\Omega_k, \nrun\}$ that come about from relaxing the strong inflation prior are true manifestations that will hold with improved data.

\subsection{$w$CDM with Massive Neutrinos}
\label{subformatwo}

\subsubsection{Enforcing the inflation prior on $\{\Omega_k,\nrun\}$}
\label{subformatwo1}

In the previous section, we considered cases with neutrinos as massless and cases with neutrinos as massive.
However, as it is well established that neutrinos are indeed massive~\cite{hirata92, davis68, Eguchi:2002dm, Ahn:2002up}, we account for the sum of neutrino masses as a free parameter in all further treatments of neutrinos. 
In Table~\ref{table:wcdm}, we explore possible degeneracies between $\{\neff, \sumnu\}$ and a constant EOS of the dark energy ($w \neq -1$). 

Beginning with ``vanilla$+w$," we constrain a constant dark energy EOS: $w = -1.10 \pm 0.11$ (as compared to $w = -1.10 \pm 0.14$ without SPT). 
Considering $w$ in conjunction with ``vanilla$+\neff$$+\sum{m_{\nu}}$," we find a reduction in
$\neff = 3.59 \pm 0.57$ (down from $\neff = 4.00 \pm 0.43$), rendering it consistent with the canonical value to within $1\sigma$. 
This is caused by the $w-\neff$ correlation discussed in Sec~\ref{introlabel} and shown in Fig.~\ref{fig:nuw}.
Expectedly, we also find a correlation between the dark energy EOS and the sum of neutrino masses (discussed in Sec.~\ref{introlabel}), the latter of which degrades by close to a factor of 2 to $\sumnu < 1.2~{\rm{eV}}$ (95\% CL).

The joint impact of $\{\neff, \sumnu\}$ on the dark energy EOS is to weaken the constraint on it by roughly a factor of 3 to $w = -1.31 \pm 0.30$.
Moreover, with the introduction of $w$, the amplitude of linear matter fluctuations is mildly shifted to smaller values at $\sigma_8 = 0.775 \pm 0.063$ (compared to $\sigma_8 = 0.798 \pm 0.053$) because of the 
anti-correlation between $w$ and $\sigma_8$ that mainly enters through the growth function (e.g.~see~\cite{Komatsu}).
The spectral index shifts further away from unity to $n_s = 0.968 \pm 0.022$ (down from $n_s = 0.987 \pm 0.013$).

%%%%%%%%%%%%%%%%%%%%%%%%%%%%%
\begin{table*}[th]\footnotesize
\begin{center}
\caption{Constraints on Cosmological Parameters using SPT+WMAP+$H_0$+BAO+SNe.}
\begin{tabular}{lcccccccccccccc|}
\hline \hline
 & & $w(a)$CDM & $w(a)$CDM & $w(a)$CDM & $w(a)$CDM\\
 & & & +\neff+$\sum{m_\nu}$ & +\neff+$\sum{m_\nu}$ & +\neff+$\sum{m_\nu}$+$Y_p$\\
 & & & & +${dn_s \over d\ln k}+\Omega_k$ & +${dn_s \over d\ln k}+\Omega_k$\\
\hline \hline
Primary & $100 \Omega_b h^2$ & $2.226 \pm 0.042$ & $2.249 \pm 0.047$ & $2.224 \pm 0.057$ & $2.163 \pm 0.078$ \cr
& $100\Omega_c h^2$ & $11.37 \pm 0.46$ & $13.3 \pm 1.1$ & $13.3 \pm 1.1$ &  $14.1 \pm 1.4$ \cr
& $10^4\theta_s$ & $104.11 \pm 0.15$ &  $103.96 \pm 0.18$ & $103.99 \pm 0.18$ & $103.66 \pm 0.36$ \cr
& $\tau$ & $0.083 \pm 0.014$ & $0.088 \pm 0.015$ & $0.089 \pm 0.016$ & $0.090 \pm 0.016$ \cr
& $n_s$ & $0.963 \pm 0.011$ & $0.978 \pm 0.015$ & $0.970 \pm 0.019$ & $0.950 \pm 0.026$ \cr
& $\ln{(10^{10} A_{s})}$ & $3.203 \pm 0.041$ & $3.194 \pm 0.044$ & $3.194 \pm 0.054$ & $3.200 \pm 0.055$ \cr
\hline
Extended & $w_0$ & $-1.10 \pm 0.17$ & $-1.05 \pm 0.19$ & $-1.08 \pm 0.21$ & $-1.12 \pm 0.21$ \cr
& $w_a$ & $0.20 \pm 0.64$ & $-0.4 \pm 1.0$ & $-0.3 \pm 1.2$ & $-0.3 \pm 1.2$ \cr
& $N_{\rm eff}$ & --- & $3.84 \pm 0.45$ & $3.57 \pm 0.59$ & $3.75 \pm 0.68$ \cr
& $\sum{m_{\nu}}~\rm{[eV]}$  & --- & $<1.2$ & $<1.4$ & $<1.8$ \cr
& ${dn_s \over d\ln k}$ & --- & --- & $-0.012 \pm 0.020$ & $-0.038 \pm 0.030$ \cr
& $100\Omega_k$ & --- & --- & $0.7 \pm 1.1$ & $1.3 \pm 1.2$ \cr
& $Y_p$ & --- & --- & --- & $0.168 \pm 0.079$ \cr
\hline
Derived & $\sigma_8$ & $0.827 \pm 0.046$ & $0.779 \pm 0.062$ & $0.763 \pm 0.076$ & $0.736 \pm 0.084$ \cr
\hline
\hline
\end{tabular}
\begin{tablenotes}
\item 
Same as Table~\ref{table:wcdmsn} but for a time-dependent parameterization of the dark energy equation of state, of the form $w(a) = w_0 + (1-a)w_a$ (as opposed to time-independent $w$).
\end{tablenotes}
\label{table:wacdmsn}
\end{center}
\end{table*}
%%%%%%%%%%%%%%%%%%%%%%%%%%%%%

\subsubsection{Relaxing the inflation prior on $\{\Omega_k,\nrun\}$}

In Table~\ref{table:wcdm}, we now consider the parameter combination ``vanilla$+\neff$$+\sum{m_{\nu}}$$+w$" in conjunction with the running of the spectral index and curvature of the universe. 
We also consider a case with the primordial helium abundance as a free parameter.

Given that we already identified separate degeneracies between $\neff-\nrun$ and $\neff-w$ in past sections,
it is not surprising that we obtain
$\neff = 3.10 \pm 0.74$ to be in even closer agreement with the canonical value for our extended parameter space. 
This is a result of the even more negative values preferred by $\nrun = -0.018 \pm 0.019$ and $w = -1.46 \pm 0.39$, shown in Figs.~\ref{fig:nuw}~and~\ref{fig:omkrun}. 
However, the upper bound on $\sumnu < 1.2~{\rm{eV}}$ is robust to the further expansion of the parameter space, such that 
this bound holds for all three cases: ``vanilla$+\neff$$+\sum{m_{\nu}}$$+w$", ``vanilla$+\neff$$+\sum{m_{\nu}}$$+\Omega_k$$+\nrun$", as well as ``vanilla$+\neff$$+\sum{m_{\nu}}$$+w$$+\Omega_k$$+\nrun$".

In addition, when allowing for $Y_p$ as an independent parameter (i.e.~considering the case ``vanilla$+\neff$$+\sum{m_{\nu}}$$+w$$+\Omega_k$$+\nrun$$+Y_p$"), the upper bound on the sum of neutrino masses is only mildly weakened, 
while the error bars are large enough that the effective number of neutrinos is consistent with values of both 3 and 4 (to within 68\% CL).
In all of the above cases we continue to find $\sigma_8$ consistent with that from cluster abundance measurements to within 68\% CL, while the spectral index is consistent with unity to within 95\% CL.

While our results are based on the construction of 3 massive neutrinos, and $(\neff - 3)$ massless degrees of freedom (as discussed in Sec.~\ref{methform}), we also considered imposing a hard prior of $\neff \geq 3$ in a new run with
``vanilla$+\neff$$+\sum{m_{\nu}}$$+w$$+\Omega_k$$+\nrun$", as this is the case where $\neff$ would be the most affected by the prior. 
Given the weak correlation between $\sumnu$ and $\neff$, the upper bound on the sum of neutrino masses doesn't change, while the data is still consistent with no extra relativistic species as 
$\neff = {{3.65}~^{3.82,~4.62}_{3.00,~3.00}}$, where the two sets of upper and lower boundaries denote 68\% and 95\% CLs, respectively. 
It is clear that our findings are qualitatively unchanged with this alternative choice of prior on the effective number of neutrinos (also see captions of Tables~\ref{table:wcdm},~\ref{table:wcdmsn},~and~\ref{table:edecdmsn}).

%%%%%%%%%%%%%%%%%%%%%%%%%%%%%
\begin{table*}[th]\footnotesize
\begin{center}
\caption{Constraints on Cosmological Parameters using SPT+WMAP+$H_0$+BAO+SNe.}
\begin{tabular}{lcccccccccccccc|}
\hline \hline
 & & eCDM & eCDM & eCDM & eCDM\\
 & & & +\neff+$\sum{m_\nu}$ & +\neff+$\sum{m_\nu}$ & +\neff+$\sum{m_\nu}$+$Y_p$\\
 & & & & +${dn_s \over d\ln k}+\Omega_k$ & +${dn_s \over d\ln k}+\Omega_k$\\
\hline \hline
Primary & $100 \Omega_b h^2$ & $2.223 \pm 0.041$ & $2.256 \pm 0.047$ & $2.196 \pm 0.058$ & $2.173 \pm 0.081$ \cr
& $100\Omega_c h^2$ & $11.56 \pm 0.42$ & $13.23 \pm 0.96$ & $13.4 \pm 1.1$ & $13.9 \pm 1.5$ \cr
& $10^4\theta_s$ & $104.00 \pm 0.17$ & $103.89 \pm 0.18$  & $103.89 \pm 0.19$ & $103.72 \pm 0.39$ \cr
& $\tau$ & $0.085 \pm 0.014$ & $0.090 \pm 0.015$ & $0.091 \pm 0.015$ & $0.094 \pm 0.016$ \cr
& $n_s$ & $0.966 \pm 0.011$ & $0.984 \pm 0.015$ & $0.966 \pm 0.019$ &  $0.957 \pm 0.027$ \cr
& $\ln{(10^{10} A_{s})}$ & $3.194 \pm 0.041$ &  $3.174 \pm 0.042$ & $3.176 \pm 0.051$ & $3.182 \pm 0.055$ \cr
\hline
Extended & $w_0$ & $-1.082 \pm 0.079$ & $-1.11 \pm 0.10$ & $-1.16 \pm 0.12$ & $-1.17 \pm 0.14$ \cr
& $\Omega_e$ & $<0.030$ & $<0.025$ & $<0.049$ & $<0.049$ \cr
& $N_{\rm eff}$ & --- & $3.85 \pm 0.43$ & $3.24 \pm 0.63$ & $3.37 \pm 0.68$ \cr
& $\sum{m_{\nu}}~\rm{[eV]}$  & --- & $<0.96$ & $<1.6$ & $<2.0$ \cr
& ${dn_s \over d\ln k}$ & --- & --- & $-0.023 \pm 0.021$ & $-0.034 \pm 0.032$ \cr
& $100\Omega_k$ & --- & --- & $1.5 \pm 1.2$ & $1.7 \pm 1.3$ \cr
& $Y_p$ & --- & --- & --- & $0.206 \pm 0.086$ \cr
\hline
Derived & $\sigma_8$ & $0.805 \pm 0.045$ & $0.773 \pm 0.063$ & $0.703 \pm 0.095$ & $0.692 \pm 0.097$ \cr
\hline
\hline
\end{tabular}
\begin{tablenotes}
\item 
Same as Table~\ref{table:wcdmsn} but for an early dark energy model with present EOS $w_0$ and density at high redshift $\Omega_e$ (as opposed to time-independent $w$).
We report the 95\% upper limit on $\Omega_e$ (and $\sumnu$ as before). 
For the ``eCDM$+\neff+\sumnu+\nrun+\Omega_k$" case, we also considered a run where we impose a hard prior of $\neff \geq 3$. Here, we find $\neff = {{3.60}~^{3.74,~4.47}_{3.00,~3.00}}$, where the two sets of upper and lower boundaries denote 68\% and 95\% CLs, respectively. The largest changes this prior induces in other parameters are seen in $n_s = 0.975 \pm 0.015$ (compared to $n_s = 0.966 \pm 0.019$), $\nrun = -0.014 \pm 0.017$ (compared to $\nrun = -0.023 \pm 0.021$),  $100\Omega_k = 1.0 \pm 1.1$ (compared to $100\Omega_k = 1.5 \pm 1.2$), and $\Omega_e < 0.042$ (as compared to $\Omega_e < 0.049$). All of the other parameters are modestly affected ($<10\%$) by our choice of prior on $\neff$.
For the vanilla$+w_0$$+\Omega_e$ case, 
we also considered a run with a hard prior $w > -1$, for which we find $\Omega_e < 0.023$ at 95\% CL (as compared to $\Omega_e < 0.019$ in Ref.~\cite{Reichardt:2011fv}).
\end{tablenotes}
\label{table:edecdmsn}
\end{center}
\end{table*}
%%%%%%%%%%%%%%%%%%%%%%%%%%%%%

%%%%%%%%%%%%%%%%%%%%%%%%%%%%%%%%%%%%%%%%%%
\subsection{$w$CDM with Massive Neutrinos, Running, and Curvature: Including Supernovae}
\label{subformathree}

Since much of the work in bringing $\neff$ in agreement with the canonical value is done by the possibility of evolving dark energy, for which the constraints from the CMB, $H_0$, and BAO measurements that we have considered are relatively weak, 
we further include SN data from the Union2 compilation in order to more effectively constrain a constant dark energy EOS and parameters with which it strongly correlates.

In Table~\ref{table:wcdmsn}, we find that the addition of SN observations help constrain the dark energy EOS to $w = -1.05 \pm 0.07$ when analyzed along with the vanilla parameters (35\% reduction in uncertainty compared to no SNe). This constraint degrades to $-1.10 \pm 0.11$ when expanding the parameter space to further include $\{\neff, \sumnu, \Omega_k, \nrun\}$, but is still a factor of 4 stronger than the equivalent case where SNe are not included in the analysis.
Improving the constraint on $w$ is helpful in breaking much of the degeneracy between dark energy and the effective number of neutrinos, resulting in $\neff = 3.88 \pm 0.44$ for the case of ``vanilla$+\neff$+$\sumnu$$+w$" (as compared to $\neff = 3.59 \pm 0.57$ without SNe, and as compared to $\neff = 4.00 \pm 0.43$ with a prior $w = -1$). However, as before, when relaxing the strong inflation prior on $\{\Omega_k, \nrun\}$ we find the effective number of neutrinos becomes consistent with the canonical value to 68\% CL (as $\neff = 3.58 \pm 0.60$).

Low-redshift SN measurements are useful in reducing the correlation between $\{\sumnu, w, H_0\}$, which drives the 1.2 eV (95\% CL) upper bound on the sum on neutrino masses for the case ``vanilla$+\neff+\sumnu$$+w$" down to 0.9 eV (Tables~\ref{table:lcdm}~and~\ref{table:wcdm}). However, since SN observations do not much improve the constraint on the curvature when added to CMB+$H_0$+BAO (shown in Fig.~\ref{fig:nuw}),
the SNe are unable to lower the upper bound on the sum of neutrino masses from 1.2 eV (95\% CL) when relaxing the strong inflation prior (i.e.~for the case ``vanilla$+\neff$$+\sum{m_{\nu}}$$+w$$+\Omega_k$$+\nrun$").
The parameter that most strongly increases the upper bound on the sum of neutrino masses when singularly added to ``vanilla+$\sumnu$" is the curvature, which renders $\sumnu < 1.0~\rm{eV}$ (95\% CL) in ``vanilla+$\sumnu$$+\Omega_k$."

Expanding the parameter space to allow $Y_p$ to vary as an independent parameter (i.e.~considering~``vanilla$+\neff$$+\sum{m_{\nu}}$$+w$$+\Omega_k$$+\nrun$$+Y_p$"), we~find a mild shift in $\neff = 3.78 \pm 0.61$ (as compared to $\neff = 3.58 \pm 0.60$), and a stronger shift in $\sumnu < 1.7~{\rm{eV}}$ (as compared to $\sumnu < 1.2~{\rm{eV}}$ at 95\% CL). Meanwhile, $Y_p = 0.176 \pm 0.079$ shows a preference for lower values but is still consistent with measurements of $Y_p$ from low-metallicity H~II regions~\cite{Peimbert:2007vm,Izotov:2007ed,Izotov:2010ca,Aver:2010wq,Aver:2010wd,Aver:2011bw}. 
For all of the non-minimal cases considered in Table~\ref{table:wcdmsn}, $n_s$ is consistent with unity to at least 95\% CL, and $\sigma_8$ mildly prefers values less than 0.8 but is still greatly consistent with cluster abundance measurements as shown in Fig.~\ref{fig:sigma8}. Moreover, we find that larger values of the dark matter density are preferred, as $\Omega_c h^2$ generally lives around $0.13 \pm 0.01$.

For the particular parameter combination that shifts $\neff$ the closest to a value of 3 from above (i.e.~``vanilla$+\neff$$+\sum{m_{\nu}}$$+w$$+\Omega_k$$+\nrun$"), we also considered a run with the prior $\neff \geq 3$ imposed. Here, we continue to find the effective number of neutrino species to be consistent with the standard value, as $\neff = {{3.74}~^{3.92,~4.68}_{3.00,~3.00}}$, where the two sets of upper and lower boundaries denote 68\% and 95\% CLs, respectively. The constraints on other parameters such as $w$ and $\sumnu$ change by less than 10\% with this alternative choice of prior.

Next, we move on to other parameterizations of the dark energy, such as the popular expansion $w(a) = w_0 + (1-a)w_a$ and an early dark energy model in which the equation of state of the dark energy tracks the equation of state of the dominant component in the universe. 

\begin{figure*}[!t]
\begin{center}
\vspace{-0.4em}
\includegraphics[bb=108.000348 229.587501 445.000979 539.410976,clip,scale=0.53]{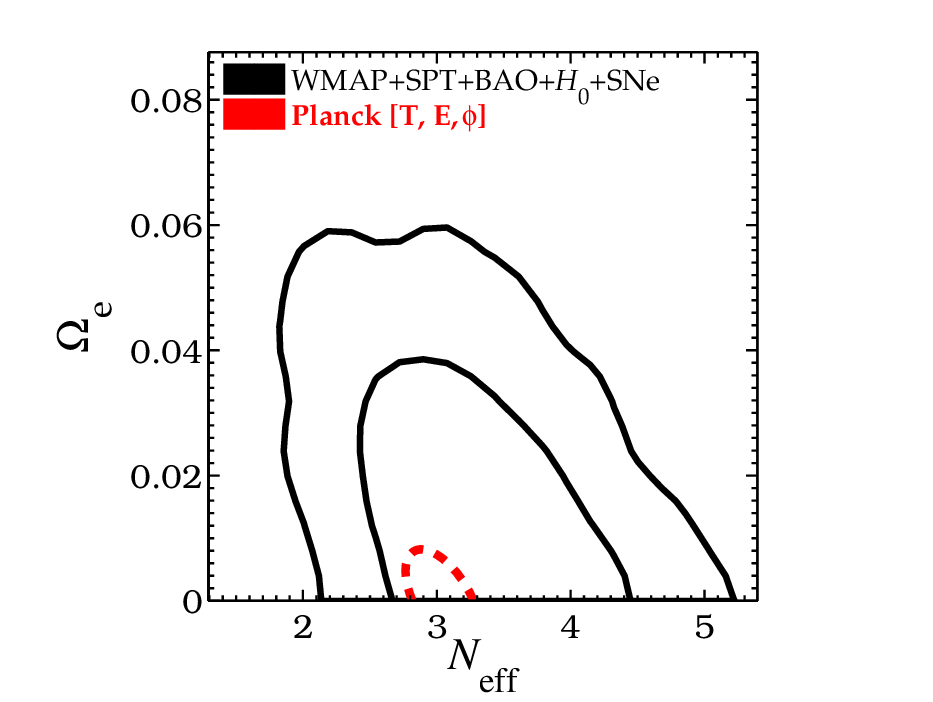}
\includegraphics[bb=177.480065 230.776344 444.334979 539.410976,clip,scale=0.53]{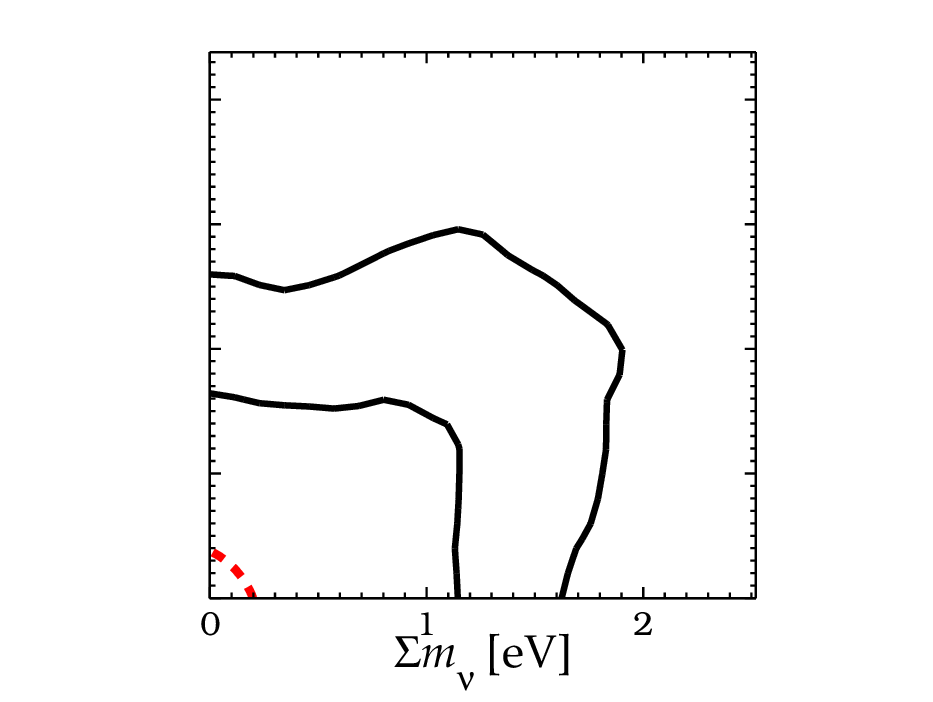}
\end{center}
\vspace{-1.9em}
\caption{Joint two-dimensional marginalized constraints on the early dark energy density $\Omega_e$ against $\{\neff, \sumnu\}$ for the extended parameter combination ``vanilla$+\neff$$+\sum{m_{\nu}}$$+w_0$$+\Omega_e$$+\Omega_k$$+\nrun$." The black confidence regions (inner 68\%, outer 95\%) are for ``WMAP+SPT+$H_0$+BAO+SNe," while the forecasted $1\sigma$ error ellipses for Planck temperature, E-mode polarization, and lensing potential power spectra (T, E, $\phi$) are shown in dashed red. Although the Fisher matrix constraints on the parameters $\{\sumnu, \Omega_e\}$ were evaluated at $\{0.17~{\rm{eV}}, 0.01\}$, they have been shifted down to $\{0, 0\}$ for simpler visual comparison with the upper bounds from present data.}
\label{fig:ede}
\end{figure*}

\subsection{Alternative Dark Energy Parameterizations}
\label{subformafive}

Given our ignorance of the nature of dark energy, once we move away from a cosmological constant, there is no adequate reason to restrict our analyses to a constant equation of state from the point of view of particle physics, in particular
if we wish to describe dark energy as a scalar field or modification of gravity~(e.g.~\cite{Frieman:2008sn,Zlatev,Huterer:2000mj,Caldwell:1997ii,Linder:2004ng}).
Thus, as an extension of the previous section, we now consider models of the dark energy in which the EOS varies with time.
While SN measurements proved useful in breaking parameter degeneracies with a constant EOS, 
we aim to understand how well these degeneracies are broken for less constrained dark energy models.

\subsubsection{Late-Time Dark Energy with Evolving \\ Equation of State}
The first of our alternative parameterizations for the dark energy is given by the two-parameter model~\cite{Linder:2002et,Linder:2004ng, Chevallier:2000qy} advocated in the report of the Dark Energy Task Force~\cite{Albrecht:2006um}:
\begin{equation}
w(a) = w_0 + (1-a)w_a,
\label{lindereqn}
\end{equation}
where $w_0$ is the EOS at present, while conventionally $w_a = -2 {dw}/{d\ln{a}}|_{a=1/2}$~\cite{Linder:2002et,Linder:2004ng}. 
Eqn.~\ref{lindereqn} may also be viewed as a first order Taylor expansion of the EOS, where $w_a = {-dw/da}|_{a=1}$.

As compared to the case with a constant dark energy EOS, we find that the new parameterization for late-time dark energy doesn't significantly change our constraints on other cosmological parameters.
Expectedly, the two most sensitive parameters are $\Omega_k$ and $\sumnu$, the constraints on which degrade by less than 15\% and 30\%, respectively.

In Table~\ref{table:wacdmsn}, we constrain $w_0 = -1.10 \pm 0.17$ and $w_a = 0.20 \pm 0.64$ for the extension ``vanilla$+w_0$$+w_a$."
When we instead use SNe from the ``Constitution" compilation~\cite{Hicken:2009dk}, we find $w_0 = -0.93 \pm 0.13$ and $w_a = -0.36 \pm 0.65$, which are consistent with the constraints on these parameters in Ref.~\cite{Komatsu:2010fb} (perfect agreement when excluding SPT). 
The difference in constraints may be traced to the larger number of SNe in the Union2 compilation (557 SNe) as compared to the Constitution compilation (397 SNe), along with the use of the SALT2 light curve fitter for the Union2 compilation as compared to the SALT fitter for the Constitution compilation.
Clearly, precise SN measurements are critical to understanding the true values of these EOS parameters.
In further extensions of our parameter space, the constraint on $w_0$ degrades by up to 20\%, while the constraint on $w_a$ degrades by up to a factor of 2.

\subsubsection{Early Dark Energy}
Late-time dark energy models suffer from the well known coincidence problem.
The value of the dark energy
density has to be fine-tuned so that it only affects the dynamics of
the universe at present. This coincidence 
problem motivates the exploration of models in which the evolution of the dark energy
density is such that it is large enough to affect the universal
dynamics even at $z>2$.

A realization of early dark energy (EDE) is given by the ``tracker"
parameterization of Doran $\&$ Robbers (2006)~\cite{DorRob}, 
where the dark energy tracks the dominant component in the
universe. In a sense, it is simpler to parameterize the dark energy
density evolution directly, rather than express it in terms of an
evolving equation of state. We use a modified form of the original
parameterization that tracks the equation
of state of the dominant energy~\cite{DorRob,Joudaki:2011nw},   
\begin{eqnarray}
\Omega_d(z) &=& \Omega_{d0}{(1+z)^{3+3w_0} \over h_w^2(z)} \nonumber\\
&+& \Omega_e v(z)\left(1-{(1+z)^{3+3w_0} \over
    h_w^2(z)}\right), \label{eq:equation2} \\
h_w^2(z)  & = & \Omega_{d0}(1+z)^{3+3w_0}+ \Omega_m (1+z)^3
\nonumber\\
&+& \Omega_r(1+z)^4 + \Omega_k (1+z)^2 ,\nonumber
\end{eqnarray}
where $\{ \Omega_r, \Omega_m\}$ are the present radiation and matter densities in units of the critical density. The
present matter density is further composed of the densities of the
cold dark matter, baryons, and massive neutrinos.
% ($\Omega_m = \Omega_c + \Omega_b + \Omega_\nu$).

As described in Ref.~\cite{Joudaki:2011nw}, the function $v(z)$ should have the properties that it
asymptotes to unity at large redshift and $v(0)=0$, thus ensuring that
$\Omega_d(z)$ asymptotes to $\Omega_e$ at large redshift and
$\Omega_d(0)=\Omega_{d0}$. We use $v(z)=1-(1+z)^{3w_0}$
\cite{DorRob}, but any other parameterization such that
$d\ln(v)/d\ln(z)={\cal O}(1)$ will give similar results. 
Note that the first term proportional to $\Omega_{d0}$ is the dark energy
density as a function of redshift for a model with present density of
dark energy $\Omega_{d0}$ and constant EOS $w_0$. Thus, in this 
parameterization with early dark energy, the effect on the dynamics of the universe at low redshift is
the same as a model with constant EOS.

\begin{table}[!t]\footnotesize
{\sc CMB Survey Properties}
\begin{tabular}{lcccccccccc|c}
\hline
Experiment & Channel & FWHM  & $\Delta T/T \times 10^6$ &  $\Delta P/T \times 10^6$ \\
Planck  & 100    & 10    & 25        &  40 \\
 & 143    & 7.1   & 16        &  30 \\
 & 217    & 5.0   & 24        &  49 \\
 \hline
\end{tabular}
\caption{Experimental specifications for the Planck mission. 
The sky fraction $f_{\rm sky} = 0.65$, and the angular multipoles extend from $\ell_{\rm min} = 2$ to $\ell_{\rm max}=2000$. The channel frequencies are given in GHz, and the angular resolutions in arcminutes.}
\label{table:cmbspecs}
\end{table}

By approximating the effect of a dark energy component with time-varying EOS using the PPF module of Ref.~\cite{Fang}, 
we allow $w_0$ to freely vary above and below the $w = -1$ boundary, unlike the treatments in Refs.~\cite{Calabrese:2011hg,Reichardt:2011fv}.
We compute the equation of state using the expression $w(z) = {-1 + {{(1+z)} \over z} {{{d \ln [\Omega_d(z) H^2(z)]} \over {3~d \ln z}}}}$,
where $H(z)$ is the Hubble parameter in a universe with radiation, matter, curvature, and dark energy (with nonzero $\Omega_e$ term)~\cite{Joudaki:2011nw}.
At z=0, $w(z)=w_0$ and increases with z, tending to 0 when the dominant component of energy density is due to pressureless matter, and to $1/3$ when the universe is dominated by radiation. 
Quantitatively, the $\Omega_e$ term (``early dark energy'') in
Eqn.~\ref{eq:equation2} constitutes $[0, 2.1, 8.0, 17.7]$\%, at
redshifts $z = [0, 1, 2, 3]$ respectively, of the overall amount of
dark energy $\Omega_d(z)$ for $w = -1$ and $\Omega_e = 0.01$.  

The impact of early dark energy on the considered observables mainly comes through increasing the expansion rate and in shifting the matter-radiation equality to a later epoch~\cite{DorRob,Joudaki:2011nw,Reichardt:2011fv,Calabrese:2011hg}. 
We can use the CMB to constrain an EDE model via its effects on $\{z_{\rm{eq}}, \theta_s, \theta_d\}$, while the improvement from BAO and SN distance measurements are modest, as the expansion rate in a model with EDE 
is designed to masquerade that of models with late-time ($z \lesssim 1$) dark energy~\cite{DorRob,Joudaki:2011nw}.
For our EDE model we fix the sound speed $c_s = 1$, while other choices have been explored in Refs.~\cite{dePutter:2010vy,Reichardt:2011fv}.

For the ``vanilla$+w_0$$+\Omega_e$" case, we find $\Omega_e < 0.030$ (95\% CL).
We also considered a run with a hard prior $w > -1$, for which we find $\Omega_e < 0.023$ at 95\% CL (as compared to $\Omega_e < 0.019$ in Ref.~\cite{Reichardt:2011fv}).
When expanding the parameter space to include the neutrino sector, we find a 20\% reduction in the upper bound on $\Omega_e$, due to its correlation with another one-tailed distribution $\sumnu$. 
While the width of the marginalized one-dimensional posterior for the EDE density parameter does widen when expanding the parameter space to include the sum of neutrino masses, the upper bound decreases because the mean of the posterior is shifted to values of the EDE density that are closer to zero.

As compared to the case where the EOS is described by a simple constant, we find modest changes in the constraints on all other parameters. However, as a result of the correlations between $\Omega_e$ and $\{\Omega_k, \nrun\}$, we find $\neff = 3.24 \pm 0.63$, $\sumnu < 1.6~{\rm{eV}}$, and $\sigma_8 = 0.703 \pm 0.095$ when relaxing the strong inflation prior (as compared to $\neff = 3.58 \pm 0.60$, $\sumnu < 1.2~{\rm{eV}}$, and $\sigma_8 = 0.774 \pm 0.072$ when $w$ is a constant). 
The upper bound on the EDE density itself degrades by a factor of 2 to $\Omega_e < 0.049$ (95\% CL). 
To obtain these shifts, the curvature and running show weak ($1\sigma$) preferences for nonzero values.

For the same parameter space, we also carried out a run with an explicit $\neff \geq 3$ prior, finding $\neff = {{3.60}~^{3.74,~4.47}_{3.00,~3.00}}$, where the two sets of upper and lower boundaries denote 68\% and 95\% CLs, respectively. 
This prior lowers the upper bound on the EDE density to $\Omega_e < 0.042$ at 95\% CL (as compared to $\Omega_e < 0.049$), 
while the constraints on other parameters such as $w$ and $\sumnu$ change by less than 10\%.
When further including $Y_p$ as a free parameter, we find qualitatively modest changes in our constraints, similar in nature to those discussed in sections~\ref{subformatwo}~and~\ref{subformathree}.

\subsection{Parameter Forecasts for Planck}
\label{subformasix}

Having discussed the present status of constraints on expanded parameter spaces with CMB, $H_0$, BAO, and SN measurements, we next explore the constraints from CMB temperature, E-mode polarization, and lensing potential power spectrum measurements with Planck~\cite{planckbb, plancksite}. To this end, we employed a Fisher matrix formalism~\cite{HuJain,Tegmark}, such that the parameter covariance matrix is given by the inverse of 
\begin{equation}
F_{\alpha\beta} = \sum_\ell \Tr \left
[{{\bf \tilde C}}_\ell^{-1} {{\partial {\bf C}_\ell} \over {\partial p_\alpha}} {{\bf \tilde C}}_\ell^{-1}
{{\partial {\bf C}_\ell} \over {\partial p_\beta}} \right ],
\label{eq:fisher}
\end{equation}
where the CMB temperature (T), E-mode polarization (E), and lensing potential ($\phi$) power spectra enter the symmetric matrix 
\begin{equation}
{\bf C_\ell} = \begin{pmatrix}
C_\ell^{\phi\phi} & C_\ell^{\phi T} & 0\\
C_\ell^{T\phi} & C_\ell^{TT} & C_\ell^{TE} \\
0 & C_\ell^{ET} & C_\ell^{EE} \\
\end{pmatrix} .
\label{eq:clall}
\end{equation}
The noise power spectra contribute additively to $\tilde{C}^{ab}(\ell) = f^{-1/2}_{{\rm sky}} \left({2 / (2\ell+1)}\right)^{1/2} \left({C^{ab}(\ell)+\delta_{ab}{N^{ab}(\ell)}}\right)$, 
where $\left\{a,b\right\} \in \left\{\phi,T,E\right\}$. 
The derivatives of the Fisher matrix are taken with respect to cosmological parameters that are defined in Table~\ref{table:priors}, and we let the ``vanilla" parameters to be given by the set $\left\{{\Omega_{b}h^2, \Omega_{c}h^2, \theta_s, \tau, n_{s}, \ln{(10^{10} A_{s})}}\right\}$.
The experimental specifications are listed in Table~\ref{table:cmbspecs}.

We take a flat $\Lambda$CDM model for the fiducial cosmology (where parameter values are based on WMAP), with $\neff = 3.04$ and $\sumnu = 0.17~{\rm{eV}}$. When EDE is included, the fiducial cosmology includes $\Omega_e = 0.01$.
We have checked that our results are not significantly affected by the choice of fiducial neutrino mass and EDE density, and expect that this holds true for the other parameters as well (e.g.~\cite{Joudaki:2011nw}).
For the terms in Eqn.~\ref{eq:fisher}, we carried out two-sided numerical derivatives
with steps of $2\%$ in most parameter values. We have confirmed the
robustness of our results to other choices of step size. 
For further details on our prescription, including how to obtain the noise power spectra, see Ref.~\cite{Joudaki:2011nw}. 
Therein, we also consider weak lensing tomography, galaxy tomography, supernovae, and extensive set of cross-correlations for future wide and deep surveys.

As shown in Figs.~\ref{fig:nuw},~\ref{fig:omkrun},~\ref{fig:ede}, Planck
will be extremely helpful in improving the constraints on extended parameter spaces.
At the $1\sigma$ level, considering the combination~``vanilla$+\neff$$+\sum{m_{\nu}}$$+w$$+\Omega_k$$+\nrun$", the effective number of neutrinos could be constrained to $\sigma(\neff) = 0.23$, mainly from the CMB temperature power spectrum, and the sum of neutrino masses to $\sigma(\sumnu) = 0.19~{\rm{eV}}$, mainly from the CMB lensing potential power spectrum (consistent with Refs.~\cite{Joudaki:2011nw,Kaplinghat:2003bh,Smith:2006nk}). These constraints on $\neff$ and $\sumnu$ are a factor of 3 stronger for both parameters than the present constraints from a joint analysis of ``WMAP+SPT+$H_0$+BAO+SN" data. 

When further allowing for the possible existence of a non-negligible component of dark energy in the high-redshift universe (i.e.~considering~``vanilla$+\neff$$+\sum{m_{\nu}}$$+w_0$$+\Omega_e$$+\Omega_k$$+\nrun$"), Planck constraints on $\sumnu$ and $\neff$ only degrade by 10\% and 20\%, respectively.
This is because the CMB temperature power spectrum achieves a factor of 6 improvement in the constraint on the EDE density at $\sigma(\Omega_e) = 0.0087$ (which only improves by 5\% in the full analysis), removing much of the degeneracy with other parameters obtained from the CMB.
Expectedly, the orientation of error ellipses for Planck and present data in Figs.~\ref{fig:nuw},~\ref{fig:omkrun},~\ref{fig:ede} match for parameters that are mainly constrained by the CMB $(T, E)$, 
while they differ for parameters, such as $w$, for which the CMB temperature and E-mode polarization provide inferior constraints. 

Moreover, we allow for the primordial helium abundance to vary in ``vanilla$+\neff$$+\sum{m_{\nu}}$$+Y_p$", where the $1\sigma$ constraints on $\{\neff,\sumnu,Y_p\}$ are at the level of $\{0.25,0.14 {\rm{eV}},0.015\}$. 
In the extended space~``vanilla$+\neff$$+\sum{m_{\nu}}$$+w$$+\Omega_k$$+\nrun$$+Y_p$," these constraints degrade to $\{0.27,0.20 {\rm{eV}},0.019\}$, respectively (such that these neutrino constraints are 
comparable with the $\Omega_e$ case).
However, the expected parameter constraints worsen significantly when considering $\{\neff, \sumnu, Y_p, \Omega_e\}$ in conjunction. 
For the maximal parameter extension (``vanilla$+\neff$$+\sum{m_{\nu}}$$+w_0$$+\Omega_e$$+\Omega_k$$+\nrun$$+Y_p$"), 
we find $1\sigma$ constraints on these four correlated parameters at the level of $\{0.51, 0.29 {\rm eV}, 0.030, 0.013\}$, respectively. In this scenario, Planck will need to be combined with external datasets to break the parameter degeneracies, in particular low-redshift measurements of the expansion history.

Thus, assuming a strong BBN prior on $Y_p$ (alternatively, at less than 10\% degradation, assuming the dark energy is a pure late-time phenomenon at the level of $\Omega_e \lsim 3\times10^{-3}$), 
it is expected that Planck alone will be able to determine the possible existence of extra relativistic species to 4$\sigma$ confidence and the sum of neutrino masses to 0.2 eV, {\it regardless} of the extent of the remaining parameter space. 
With the ability to strongly constrain $\sumnu$, there is promise for Planck to find evidence for nonzero neutrino mass, in particular when combined with probes of the large-scale structure~\cite{Joudaki:2011nw} (also see~\cite{Cooray:1999rv,Abazajian:2002ck,Lesgourgues:2004ps,Hannestad,Hannestad:2002cn,Hu:1997mj,Eisenstein:1998hr,Joudaki:2009re,Carbone:2010ik,Carbone:2011by}).

\section{Conclusions}
\label{conclusionlabel}

With the latest cosmological data sets of the cosmic microwave background (WMAP7+SPT), baryon acoustic oscillations (SDSS+2dF), supernovae (Union2), and the Hubble constant (HST), 
we have explored in closer detail the dependence of constraints on the effective number of neutrino species and the sum of neutrino masses on our assumptions of other cosmological parameters, including the curvature of the universe, running of the spectral index, primordial helium abundance, evolving late-time dark energy, and early dark energy. 

In a combined analysis of the effective number of neutrinos and sum of neutrino masses (with 6 other $\Lambda$CDM parameters), we find mild (2.2$\sigma$) preference for additional light degrees of freedom. However, the effective number of neutrinos is consistent with three massive neutrinos and no extra relativistic species to $1\sigma$ when further including $\{w, \Omega_k, \nrun\}$ as free parameters. The transformation of a constant EOS to one that varies with time ($w_0, w_a$) doesn't significantly
change the constraints on $\neff$ and $\sumnu$ (less than 10\% and 30\%, respectively). The agreement with $\neff = 3.046$ improves with the possibility of an early dark energy component, itself constrained to be less than 5\% of the critical density (95\% CL) in our maximally expanded parameter space. 

Added to a minimal $\Lambda$CDM universe, $\sumnu < 0.45$~eV (95\% CL), dominantly from WMAP+HST.
The sum of neutrino masses is bounded at 1.2~eV (95\% CL) when jointly allowing for a constant dark energy equation of state, curvature, and running to vary as free parameters. 
The upper bound degrades to 2.0~eV (95\% CL) when further including the primordial helium abundance and early dark energy density as additional degrees of freedom.
The single parameter that most strongly increases the upper bound on the sum of neutrino masses when added to ``vanilla+$\sumnu$" is the curvature of the universe, which weakens the bound by more than a factor of 2 to $\sumnu < 1.0~\rm{eV}$ (95\% CL) in ``vanilla+$\sumnu$$+\Omega_k$."

In extensions of the standard cosmological model that minimally allow for nonzero neutrino masses and additional light degrees of freedom, the derived amplitude of linear matter fluctuations $\sigma_8$ is found consistent with low-redshift cluster abundance measurements to within $1\sigma$, 
and the spectral index agrees with unity to within 1 to 2 $\sigma$.
Moreover, larger values of the dark matter density are preferred, as $\Omega_c h^2$ generally lives around $0.13 \pm 0.01$.
When allowing the primordial helium abundance to vary as a free parameter, we consistently find $1\sigma$ agreements with estimates from observations of low-metallicity extragalactic H~II~regions.

With the advent of increasingly sensitive CMB and large-scale structure data~\cite{plancksite,planckbb,sdsssite,dessite,euclidsite,lsstsite,Baumann:2008aq,Bock:2009xw}, our understanding of the neutrino sector depends critically on the ability to distinguish its signatures from other cosmological parameters. 
Fortunately, even for extended parameter spaces, Planck alone could determine the possible existence of extra relativistic species at the $4\sigma$ level and constrain the sum of neutrino masses to 0.2 eV (68\% CL).
Next-generation probes of the expansion history and large-scale structure hold the key to further improving these estimates.

%%%%%%%%%%%%%%%%%%%%%%%%%%%%%%%%%%%%%%%%%%%%%%%%%%%%%%%%%%%%%%%%%%%%%%%%%
\smallskip
{\it Acknowledgements:} 
We thank Manoj Kaplinghat for valuable discussions and feedback throughout this work.
We also thank Michael Mortonson for help with a bug in the PPF module when implemented in CosmoMC.
We much appreciate useful discussions with John Beacom, Francesco De Bernardis, Zhen Hou, Lloyd Knox, Gregory Martinez, Jose O\~norbe, Joseph Smidt, and Gary Steigman.
We acknowledge the use of CAMB and CosmoMC packages~\cite{LCL,Lewis:2002ah}, and  support from the US Dept.~of Education through GAANN at UCI.
%%%%%%%%%%%%%%%%%%%%%%%%%%%%%%%%%%%%%%%%%%%%%%%%%%%%%%%%%%%%%%%%%%%%%%%%%

\end{document}